\documentclass[10pt,twocolumn,letterpaper]{article}

\usepackage[pagenumbers]{cvpr} 
\usepackage{booktabs}
\usepackage{siunitx}
\usepackage{makecell}
\usepackage{multirow}
\usepackage{caption} 
\usepackage{graphicx} 
\usepackage{booktabs}   
\usepackage{pifont}     
\usepackage{multirow}   

\usepackage{listings}
\usepackage{algorithm}
\usepackage{algpseudocode}
\usepackage{amsmath}
\definecolor{cvprblue}{rgb}{0.21,0.49,0.74}
\usepackage{booktabs}
\usepackage{multirow}
\usepackage{graphicx}
\usepackage[table]{xcolor}
\usepackage{marvosym}
\definecolor{resultred}{RGB}{210,0,0}
\definecolor{resultblue}{RGB}{45,105,255}

\newcommand{\rednum}[1]{\textcolor{resultred}{\textbf{#1}}}
\newcommand{\bluenum}[1]{\textcolor{resultblue}{\textbf{#1}}}
\usepackage[pagebackref,breaklinks,colorlinks,allcolors=cvprblue]{hyperref}

\definecolor{tablegray}{RGB}{242,242,242}

\def\confName{CVPR}
\def\confYear{2027}

\title{Dem2TA: Beyond Pixels, Towards Perception}

\author{
Ping Chen$^{1}$ \hspace{0.3cm}
Zhongwei Lin$^{1}$ \hspace{0.3cm} 
Xiang He$^{1}$ \hspace{0.3cm}
Xiangming Wang$^{1}$ \hspace{0.3cm}
Yongyong Chen$^{1}$ \hspace{0.3cm} \\[0.4em]
Kai Zhang$^{3}$ \hspace{0.3cm}
Jiezhang Cao$^{2}$ \hspace{0.3cm}
Jingyong Su$^{1}$ \hspace{0.3cm}
Jie Liu$^{1}$ \hspace{0.3cm}
Haijin Zeng$^{1}$\textsuperscript{\Letter}
\thanks{\Letter\ Corresponding author.} \\[0.7em]
{\normalfont $^{1}$Harbin Institute of Technology, Shenzhen} \\
{\normalfont $^{2}$Shanghai Jiao Tong University} \\
{\normalfont $^{3}$Nanjing University (Suzhou)} \\[0.3em]
{\normalfont \texttt{pingchen2004@gmail.com},
\texttt{haijin.zeng2018@gmail.com}} \\
{\normalfont Code: \url{https://github.com/cipi666/Dem2TA}}
}

\begin{document}
\maketitle


\begin{abstract}

Task-oriented image signal processors (ISPs) adapt RAW-to-RGB processing for downstream perception. However, explicit demosaicing is often simplified, kept fixed, or optimized primarily for reconstruction quality, leaving this stage insufficiently integrated into task adaptation. We propose \textbf{Dem2TA}, a framework that incorporates fidelity-supervised demosaicing into the adaptation of the perception pipeline. Starting from a pretrained demosaicing model, we jointly adapt the reconstruction frontend and downstream network using reconstruction and task supervision, together with complementary semantic and task representation alignment. We then freeze the adapted frontend and jointly optimize a lightweight local ISP with the downstream model to refine task-specific rendering without changing the reconstructed linear RGB. Experiments on synthetic RAW detection and box-prompted segmentation demonstrate the effectiveness of representation-guided joint adaptation and subsequent local rendering. Evaluation on a manually verified real-capture dataset further supports transfer beyond synthetic inputs. Compared with TA-ISP on the same adapted foundation, Dem2TA improves detection AP by 0.46 points and segmentation IoU by 1.06 points on synthetic COCO RAW, and segmentation IoU by 0.63 points on real captures.

\end{abstract}
    
\section{Introduction}
\label{sec:intro}

An image signal processor (ISP)~\cite{DBLP:journals/tip/SchwartzGB19} transforms RAW sensor measurements into RGB images through operations such as demosaicing~\cite{DBLP:conf/icassp/MalvarHC04,DBLP:conf/iccv/LeePJKJRC23}, color correction~\cite{DBLP:conf/imaging/FinlaysonGF16,DBLP:conf/cvpr/LiTT26}, and tone mapping~\cite{DBLP:journals/tog/GharbiCBHD17}. These operations are essential to everyday photography and also determine the visual inputs available to detection and segmentation systems. Conventional ISPs are primarily designed for human viewing, whereas machine perception places different demands on the processed images. This distinction has motivated task-driven ISPs~\cite{DBLP:conf/nips/WangXFXG24,DBLP:conf/eccv/CuiH24,DBLP:journals/pami/CuiYH26,DBLP:conf/iccv/GuoGYLP25,DBLP:conf/cvpr/ChenXZSG26}, which learn image transformations or processing parameters under downstream supervision rather than optimizing visual quality alone.

\begin{figure}[t]
    \centering
    \includegraphics[width=0.49\textwidth]{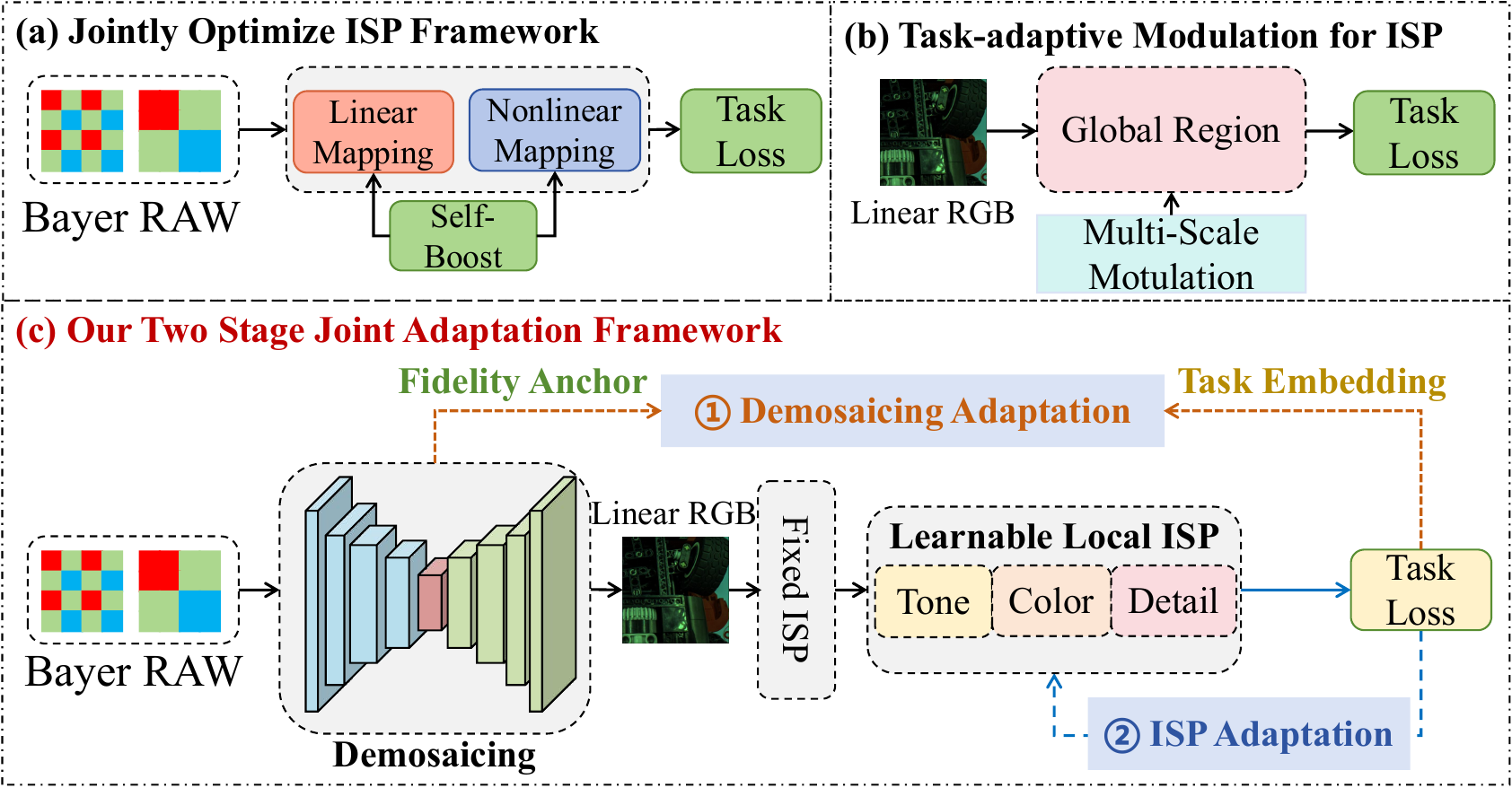}
    \caption{\textbf{Comparison of task-driven ISP paradigms.} (a,b) Joint image-mapping optimization and task-adaptive modulation. (c) Our two-stage approach jointly adapts reconstruction and perception
    through complementary supervision, then optimizes a local ISP with the reconstruction frontend frozen.
}
    \label{fig:motivation}
\end{figure}

Task-driven ISP methods use downstream feedback to learn RAW-to-RGB mappings or to control processing parameters, as illustrated in Fig.~\ref{fig:motivation}(a,b). Dark-ISP learns linear and nonlinear mappings from packed Bayer data~\cite{DBLP:conf/iccv/GuoGYLP25}, where the four Bayer samples are reduced to RGB at the packed resolution by averaging the two green samples; this does not explicitly reconstruct a full-resolution linear-RGB image. AdaptiveISP selects and configures processing operations~\cite{DBLP:conf/nips/WangXFXG24}, while TA-ISP predicts spatial modulation at multiple scales~\cite{DBLP:conf/cvpr/ChenXZSG26}. RAW-Adapter connects RAW processing to pretrained perception models through adapters~\cite{DBLP:conf/eccv/CuiH24,DBLP:journals/pami/CuiYH26}. These approaches show that task feedback can improve the rendered image, but they do not jointly adapt a fidelity-supervised demosaicing model and the downstream encoder through representation alignment. When a separate demosaicing frontend is used, it is commonly fixed or trained for image quality before task optimization. Thus, the reconstruction stage itself remains an underused part of task adaptation.

We therefore ask how demosaicing can contribute to perception beyond producing an accurate linear-RGB image. A reconstruction with higher pixel fidelity need not provide the most useful input for every downstream ISP and task, as our experiments later show. The challenge is to make the reconstruction process responsive to object-level supervision while retaining its fidelity, rather than expecting subsequent rendering to recover all task-relevant information. This also shapes the ISP design, where basic white balance~\cite{DBLP:conf/iccv/FinlaysonDF93}, color correction, and gamma mapping~\cite{DBLP:conf/imaging/AndersonMCS96} provide a conventional rendering, while the remaining task-dependent changes may vary across regions and object boundaries. We keep these basic operations fixed and learn lightweight local adjustments on top of the adapted demosaicing output. Separating the two stages lets local ISP optimization build on a stable reconstruction input, as illustrated in Fig.~\ref{fig:motivation}(c).

To this end, we propose \textbf{Dem2TA}, a framework that brings explicit, fidelity-supervised demosaicing into task-oriented ISP adaptation, illustrated in Fig.~\ref{fig:motivation}(c). Our key insight is to make the reconstruction stage participate in the adaptation of the entire perception pipeline while retaining reconstruction supervision. Starting from a pretrained demosaicing model, we jointly adapt the reconstruction frontend and downstream network under reconstruction supervision. To guide this adaptation beyond pixel-level fidelity, we introduce complementary representation alignment, where frozen DINO~\cite{DBLP:journals/tmlr/OquabDMVSKFHMEA24} features provide a semantic reference, while task-encoder features supply guidance tied to the downstream objective. We then freeze the adapted reconstruction frontend and jointly optimize a lightweight local ISP with the downstream model, enabling further task-specific rendering.

Our contributions are summarized as follows:
\begin{itemize}
    \item We propose Dem2TA to address the disconnect between reconstruction-oriented demosaicing and task adaptation, bringing explicit reconstruction into joint frontend--downstream optimization.
    \item We use complementary semantic and task representation alignment to guide this joint adaptation, followed by lightweight local ISP optimization on the fixed, adapted reconstruction.
    \item We evaluate reconstruction fidelity and downstream performance together on synthetic RAW data and a manually verified real-capture evaluation set, revealing both the benefits and configuration-dependent trade-offs of the proposed adaptation.
\end{itemize}

\section{Related Works}

\subsection{Image Signal Processing for Visual Perception}

Early learning-based ISP methods, such as Learning to See in the Dark~\cite{DBLP:conf/cvpr/ChenCXK18} and DeepISP~\cite{DBLP:journals/tip/SchwartzGB19}, explored neural RAW-to-RGB processing primarily for image quality. Subsequent research shifted toward optimizing image processing for downstream perception. ReconfigISP~\cite{DBLP:conf/iccv/YuLPLG21} searched for task-dependent processing architectures and parameters, while GenISP~\cite{DBLP:conf/cvpr/MorawskiCLDHH22} learned a compact neural ISP under detector supervision. IA-YOLO~\cite{DBLP:conf/aaai/0005RYGZ022} similarly predicted differentiable image-processing parameters and jointly trained the processing module with an object detector.

Later methods introduced more adaptive control over image processing. GDIP~\cite{DBLP:conf/icra/KalwarPAKGK23} combined parallel processing operations through learned gates, whereas DynamicISP~\cite{DBLP:conf/iccv/YoshimuraOIO23} used recognition feedback to control ISP parameters. AdaptiveISP~\cite{DBLP:conf/nips/WangXFXG24} extended this direction by learning both operation sequences and their parameters through reinforcement learning. In parallel, RAW-Adapter~\cite{DBLP:conf/eccv/CuiH24,DBLP:journals/pami/CuiYH26} introduced input- and model-level adapters to connect ISP processing with pretrained perception networks. More recently, Dark-ISP~\cite{DBLP:conf/iccv/GuoGYLP25} learned adaptive linear mappings and polynomial nonlinear transformations, and TA-ISP~\cite{DBLP:conf/cvpr/ChenXZSG26} introduced lightweight modulation at global, regional, and pixel scales. Our approach jointly adapts a pretrained demosaicing frontend and the downstream network before optimizing a local ISP, explicitly separating reconstruction adaptation from subsequent rendering optimization.

\subsection{Representation Learning for Visual Perception}

\begin{figure*}[t]
    \centering
    \includegraphics[width=\textwidth]{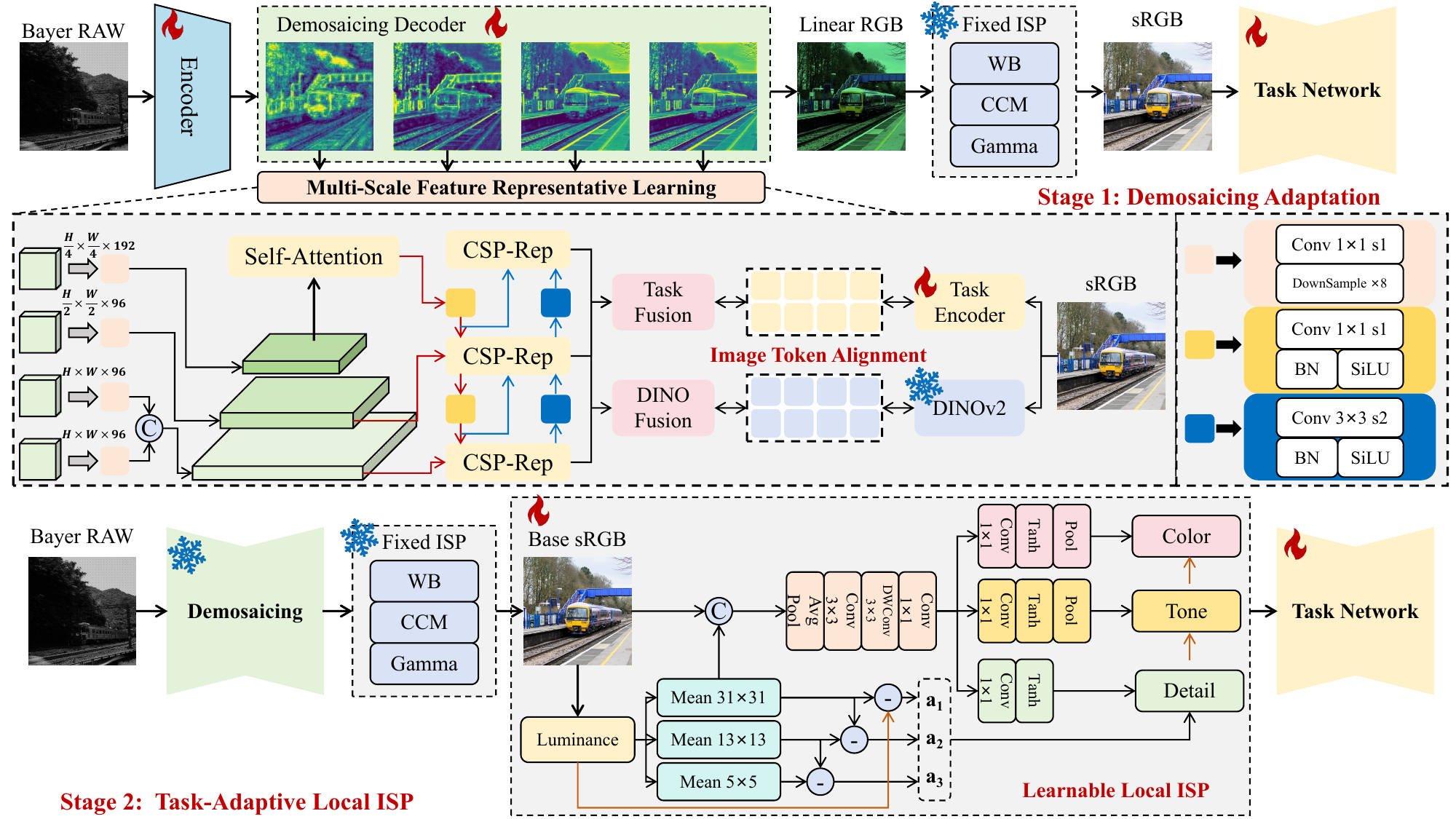}
    \caption{Overview of Dem2TA, which adapts fidelity features for perception before refining task rendering. Stage~1 jointly adapts the demosaicing frontend and downstream model with fidelity and complementary representation supervision. Stage~2 freezes the adapted frontend and jointly optimizes a local ISP with the downstream model. Flames and snowflakes indicate trainable and frozen modules.}
    \label{fig:overview}
\end{figure*}

Pretrained visual representations provide transferable information beyond pixel-level supervision. DINO~\cite{DBLP:conf/iccv/CaronTMJMBJ21} demonstrates the emergence of object-related structure through self-supervised learning, and DINOv2~\cite{DBLP:journals/tmlr/OquabDMVSKFHMEA24} develops robust representations transferable across visual tasks. MAE~\cite{DBLP:conf/cvpr/HeCXLDG22} learns representations through masked image reconstruction, while I-JEPA~\cite{DBLP:conf/cvpr/AssranDMBVRLB23} predicts missing content in a learned representation space. These approaches offer different ways to connect image structure with semantic understanding. Task-oriented foundation models provide complementary supervision: SAM~\cite{DBLP:conf/iccv/KirillovMRMRGXW23} learns spatial representations for promptable segmentation, and MobileSAM~\cite{DBLP:journals/corr/abs-2306-14289,DBLP:conf/aaai/ShuLTZCL0025} transfers its image embeddings to a lightweight encoder.

Semantic representations have also been used to guide image reconstruction. More recently, semantic-prior distillation~\cite{DBLP:conf/cvpr/ZhangL0CL0XY024} has introduced knowledge from segmentation models such as SAM~\cite{DBLP:conf/iccv/KirillovMRMRGXW23} into lightweight image restoration networks. REPA~\cite{DBLP:conf/iclr/YuKJJHSX25} aligns intermediate generative features with pretrained visual representations, demonstrating the value of semantic supervision for learning image generation. Related work on joint distillation and mutual learning~\cite{DBLP:conf/cvpr/RanzingerHK024,DBLP:conf/cvpr/WangVFVFMRTP22,DBLP:conf/cvpr/ZhangXHL18} further explores how multiple sources of knowledge can support representation learning.

We use complementary representation alignment to support joint adaptation of the reconstruction frontend and downstream encoder, allowing semantic supervision and reconstruction priors to contribute to both components.

\section{Methods}
\label{sec:method}

\subsection{Problem Formulation and Overview}
\label{sec:overview}

Given a Bayer RAW image $\mathbf{x}$, we aim to preserve reconstruction fidelity while improving downstream perception. Our pipeline consists of a demosaicing network $D_{\theta}$, a fixed ISP $\mathcal{P}$, a learnable local ISP $A_{\phi}$, and a task network $T_{\psi}$:
\begin{equation}
    \mathbf{z}=D_{\theta}(\mathbf{x}), \qquad
    \hat{\mathbf{y}}=T_{\psi}\bigl(A_{\phi}(\mathcal{P}(\mathbf{z}))\bigr),
    \label{eq:pipeline}
\end{equation}
where $\mathbf{z}$ is the reconstructed linear RGB image and $\hat{\mathbf{y}}$ denotes the task prediction. The fixed ISP converts linear RGB to sRGB through white balance, color correction, and gamma mapping.

As shown in Fig.~\ref{fig:overview}, Dem2TA adapts this pipeline in two stages. Stage~1 starts from a pretrained demosaicing model and jointly adapts the frontend and task network. Reconstruction supervision constrains image fidelity, while complementary DINO and task-embedding alignment guides the multiscale decoder features toward object-level organization. The local ISP is bypassed during this stage. Stage~2 freezes the adapted frontend and jointly trains the local ISP with the task network, enabling spatially adaptive rendering on a stable reconstruction foundation.

\subsection{Demosaicing Adaptation}
\label{sec:demosaicing_adaptation}

\begin{figure}[t]
    \centering
    \includegraphics[width=\linewidth]{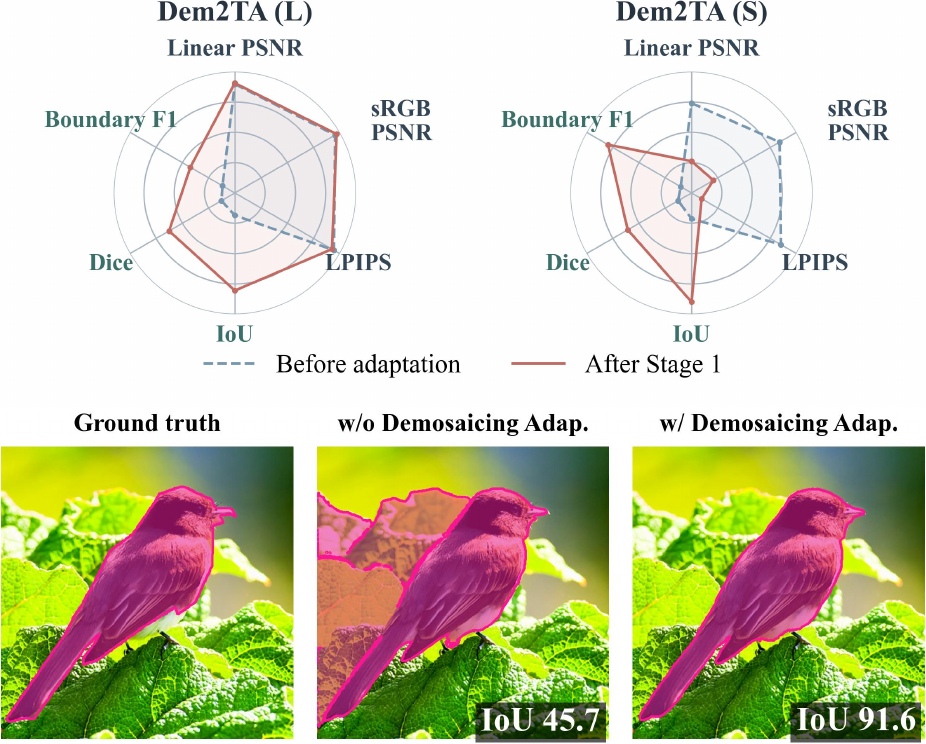}%
    \caption{Motivation for demosaicing adaptation. (a) Reconstruction oriented training optimizes demosaicing for fidelity independently of task learning. (b) Dem2TA combines reconstruction supervision, complementary representation alignment, and task feedback to jointly adapt the frontend and downstream network.}
    \label{fig:adaptation_motivation}
\end{figure}

Reconstruction-oriented demosaicing learns to recover missing color samples, but its fidelity objective does not account for how the downstream model uses the reconstructed image. Training the two components separately therefore leaves their interaction unoptimized. Our goal is to adapt them together: task feedback guides reconstruction, while the downstream encoder learns to use the adapted inputs under a reconstruction constraint (Fig.~\ref{fig:adaptation_motivation}). We initialize the frontend with a pretrained Restormer-based demosaicing model~\cite{DBLP:journals/corr/abs-2608-07916,DBLP:conf/cvpr/ZamirA0HK022}, and introduce complementary representation alignment to connect its reconstruction hierarchy with object-level structure.

\subsubsection{Multi-scale feature fusion.}
\label{sec:feature_fusion}
The decoder reconstructs the image progressively, where coarse features provide spatial context, while finer features retain local structure. We aggregate all four decoder levels so that representation supervision reaches both types of information. Each feature is channel-projected with a $1\times1$ convolution and spatially downsampled. The two finest-resolution features are concatenated and projected into one level, forming a three-level pyramid.

Self-attention is applied at the coarsest level to capture long-range context at reduced spatial cost. Bidirectional convolutional fusion~\cite{DBLP:conf/cvpr/ZhaoLXWWDLC24} then combines the pyramid: the top-down path upsamples coarse features and concatenates them with finer features, while the bottom-up path uses stride-2 convolutions to propagate local information back to coarser levels. CSP-Rep blocks~\cite{DBLP:conf/cvpr/ZhaoLXWWDLC24} integrate the concatenated features while retaining a shortcut pathway. Two separate projection heads map the fused pyramid into the DINO and task-encoder spaces. This lets semantic and task supervision shape multiple reconstruction scales without replacing the linear-RGB reconstruction branch. Block-level details are given in supplement~\ref{sec:supp_fusion_details}.

\subsubsection{Complementary alignment and joint adaptation.}
\label{sec:representation_alignment}
\begin{figure}[htbp]
    \centering
    \includegraphics[width=\linewidth]{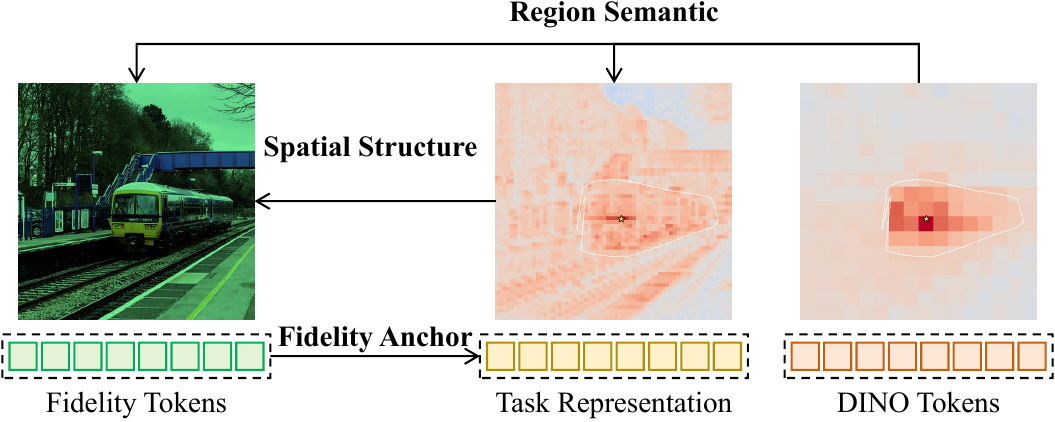}
    \caption{Motivation of complementary representation alignment. The pretrained demosaicing model provides a fidelity anchor, the task encoder supplies task-related spatial structure, and DINO provides regional semantics to bridge reconstruction and perception. These complementary priors guide the fused multiscale decoder features during joint adaptation.}
    \label{fig:alignment}
\end{figure}

Reconstruction supervision anchors the image output, but does not explicitly organize decoder features around objects. DINO provides a shared regional semantic reference, while task-encoder features supply spatial cues tied to the downstream objective (Fig.~\ref{fig:alignment}). Aligning the fused features with both references connects reconstruction detail to semantic structure without requiring the two reference spaces to match directly. We align the projected features with sRGB reference embeddings using Smooth-$L_1$ loss for DINO and MSE for task features. The reconstructed image passes through the fixed ISP to the downstream network, and task feedback jointly adapts both models. Encoder settings are detailed in Sec.~\ref{sec:supp_alignment_settings}. We optimize
\begin{equation}
    \mathcal{L}_{\mathrm{adapt}}
    =
    \lambda_{\mathrm{rec}}\mathcal{L}_{\mathrm{rec}}
    +
    \lambda_d\mathcal{L}_{\mathrm{DINO}}
    +
    \lambda_e\mathcal{L}_{\mathrm{emb}}
    +
    \lambda_{\mathrm{task}}\mathcal{L}_{\mathrm{task}},
    \label{eq:adaptation_objective}
\end{equation}
where the four terms supervise reconstruction, semantic alignment, task alignment, and task predictions, respectively. Thus, alignment shapes the reconstruction features while joint task training adapts how they are used through the rendered image. The local ISP is bypassed in this stage, and the auxiliary alignment branches are removed at inference. Implementation details are provided in supplement~\ref{sec:supp_fusion_details}.

\begin{table*}[t]
    \centering
    \caption{Comparison on COCO detection and segmentation, grouped by processing pipeline. AdaptiveISP and TA-ISP are each evaluated with two demosaicing frontends. The best and second-best results are highlighted in red and blue, respectively.}
    \label{tab:coco_comparison}
    \setlength{\tabcolsep}{4pt}
    \renewcommand{\arraystretch}{1.12}

    \resizebox{\textwidth}{!}{%
    \begin{tabular}{ll*{12}{c}}
        \toprule
        \multirow{2}{*}{Method}
        & \multirow{2}{*}{Reference}
        & \multicolumn{6}{c}{Detection (COCO)}
        & \multicolumn{6}{c}{Segmentation (COCO)} \\
        \cmidrule(lr){3-8}
        \cmidrule(lr){9-14}
        &
        & AP$\uparrow$
        & AP$_{50}\uparrow$
        & AP$_{75}\uparrow$
        & AP$_S\uparrow$
        & AP$_M\uparrow$
        & AP$_L\uparrow$
        & IoU$\uparrow$
        & Dice$\uparrow$
        & Prec.$\uparrow$
        & Recall$\uparrow$
        & Boundary F1$\uparrow$
        & Boundary IoU$\uparrow$ \\
        \midrule

        \rowcolor{gray!15}
        \multicolumn{14}{l}{
            \textit{Reconstruction-based pipelines with joint task fine-tuning}
        } \\
        \addlinespace[2pt]

        JD3Net
        & ICCP 2025
        & 26.439 & 40.235 & 28.038 & 7.878 & 27.705 & 44.804
        & 76.503 & 85.232 & 86.569 & 86.469 & 65.005 & 25.373 \\

        JD3Net + Fixed ISP
        & ICCP 2025
        & 31.289 & 47.655 & 33.915 & 9.489 & 34.251 & 49.569
        & 76.913 & 85.434 & 87.527 & 85.818 & 66.580 & 26.177 \\

        \midrule
        \rowcolor{gray!15}
        \multicolumn{14}{l}{
            \textit{Mosaic-input methods without a standalone demosaicing network}
        } \\
        \addlinespace[2pt]

        RAW-Adapter
        & ECCV 2024
        & 26.785 & 41.065 & 28.065 & 7.764 & 27.938 & 44.928
        & 79.034 & 87.190 & 86.248 & \bluenum{89.903}
        & 67.798 & 25.402 \\

        Dark-ISP
        & ICCV 2025
        & 28.205 & 43.373 & 29.659 & 8.353 & 29.984 & 46.542
        & 79.372 & 87.435 & 87.377 & 89.094 & 68.665 & 25.970 \\

        \midrule
        \rowcolor{gray!15}
        \multicolumn{14}{l}{
            \textit{ISP comparison with different demosaicing frontends}
        } \\
        \addlinespace[2pt]

        JD3Net + AdaptiveISP
        & NeurIPS 2024
        & 27.237 & 41.685 & 28.554 & 7.642 & 28.709 & 45.408
        & 78.886 & 87.067 & 86.598 & 89.258 & 67.928 & 25.164 \\

        Dem2TA (L) + AdaptiveISP
        & NeurIPS 2024
        & 30.482 & 47.251 & 33.293 & 9.323 & 33.912 & 49.569
        & 77.021 & 85.392 & 87.432 & 85.973 & 66.599 & 26.213 \\

        \addlinespace[3pt]

        JD3Net + TA-ISP
        & CVPR 2026
        & 27.023 & 41.542 & 28.359 & 7.766 & 28.277 & 45.048
        & 79.088 & 87.213 & 86.794 & 89.343 & 68.165 & 25.432 \\

        Dem2TA (L) + TA-ISP
        & CVPR 2026
        & 32.282 & 48.548 & 34.320 & 10.760 & 34.750 & 50.434
        & 80.096 & 87.942 & 89.431 & \rednum{89.941}
        & 69.719 & 26.670 \\

        \midrule
        \rowcolor{gray!15}
        \multicolumn{14}{l}{\textit{Our complete framework}} \\
        \addlinespace[2pt]

        \textbf{Dem2TA (L)}
        & Ours
        & \bluenum{32.746}
        & \bluenum{48.971}
        & \bluenum{34.830}
        & \bluenum{11.785}
        & \rednum{35.188}
        & \bluenum{50.859}
        & \rednum{81.152}
        & \rednum{88.640}
        & \rednum{90.114}
        & 88.563
        & \rednum{71.570}
        & \rednum{28.153} \\

        \textbf{Dem2TA (S)}
        & Ours
        & \rednum{32.757}
        & \rednum{49.183}
        & \rednum{35.298}
        & \rednum{12.236}
        & \bluenum{34.964}
        & \rednum{51.160}
        & \bluenum{80.846}
        & \bluenum{88.438}
        & \bluenum{89.667}
        & 88.679
        & \bluenum{70.971}
        & \bluenum{28.007} \\

        \bottomrule
    \end{tabular}%
    }
\end{table*}

\subsection{Local ISP Adaptation}
\label{sec:local_isp}

After demosaicing adaptation, we further adjust ISP to improve downstream task performance. Studies of ISP tuning show that detection models respond differently to changes in brightness, contrast, and color~\cite{DBLP:journals/jimaging/MolloyDMWHEDJG23}, motivating task-guided adjustments rather than uniform enhancement. We freeze the demosaicing frontend and jointly train a local ISP with the downstream model. Its three branches control tone, color, and detail, covering complementary appearance properties explored in task-driven processing~\cite{DBLP:conf/cvpr/MorawskiCLDHH22,DBLP:conf/icra/KalwarPAKGK23}. 

Given the base image $\mathbf{b}=\mathcal{P}(D_{\theta^\star}(\mathbf{x}))$, we compute the luminance proxy $Y=\mathbf{w}^{\top}\mathbf{b}$, where $\mathbf{w}=(0.2126,0.7152,0.0722)^{\top}$. As shown in Fig.~\ref{fig:overview}, a shared controller predicts bounded coefficient maps from $\mathbf{b}$. Here, $Y$ is a brightness proxy in rendered RGB. Tone and color maps are pooled and upsampled for smooth regional changes, while detail maps retain finer spatial variation.

\paragraph{Local tone and color adjustment.}
Tone controls regional brightness, while color provides a separate means of adjusting chromatic differences:
\begin{equation}
    \begin{aligned}
        \Delta_{\mathrm{tone}}(p)
        &= \alpha_t\,t(p)\,g(Y(p))\,\mathbf{1},\\
        \Delta_{\mathrm{color}}(p)
        &= \alpha_c\,g(Y(p))
        \begin{bmatrix}
            u(p)\\
            -\dfrac{w_Ru(p)+w_Bv(p)}{w_G}\\
            v(p)
        \end{bmatrix},
    \end{aligned}
    \label{eq:tone_color}
\end{equation}
where $\alpha_t$ and $\alpha_c$ set the tone and color adjustment scales, $p$ indexes pixels, $t,u,v\in[-1,1]$ are predicted coefficients, and $\mathbf{1}=(1,1,1)^{\top}$. The factor $g(Y)=Y(1-Y)$ reduces changes near intensity endpoints. The tone residual is shared across channels, whereas the color residual satisfies $\mathbf{w}^{\top}\Delta_{\mathrm{color}}=0$ before clipping.

\paragraph{Multi-scale detail adjustment.}
A uniform sharpening strength cannot independently emphasize weak boundaries and suppress distracting texture. We instead predict local weights for three detail bands:
\begin{equation}
    \begin{aligned}
        &D_1 = Y-\mathcal{M}_{5}(Y), D_2 = \mathcal{M}_{5}(Y)-\mathcal{M}_{13}(Y),\\
        &D_3 = \mathcal{M}_{13}(Y)-\mathcal{M}_{31}(Y),\\
        &\Delta_{\mathrm{detail}}(p) = \alpha_d\sum_{i=1}^{3}a_i(p)D_i(p)\,\mathbf{1},
    \end{aligned}
    \label{eq:local_detail}
\end{equation}
where $\mathcal{M}_{k}$ is a $k\times k$ mean filter, $a_i\in[-1,1]$, and $\alpha_d$ sets the detail adjustment scale. Signed weights allow enhancement or suppression at each scale. The residual is shared across RGB channels to avoid color-channel sharpening.

\paragraph{Task-guided optimization.}
The three branches predict complementary residuals from the base image:
\begin{equation}
    \mathbf{I}
    =
    \operatorname{clip}_{[0,1]}
    \left(
        \mathbf{b}
        +\Delta_{\mathrm{tone}}
        +\Delta_{\mathrm{color}}
        +\Delta_{\mathrm{detail}}
    \right).
    \label{eq:local_output}
\end{equation}
Zero-initialized prediction heads make all three residuals zero, preserving the base image at initialization. We jointly optimize the local ISP and downstream model with task supervision and appearance regularization:
\begin{equation}
\begin{aligned}
    \min_{\phi,\psi}\quad
    &\mathcal{L}_{\mathrm{task}}\left(T_{\psi}(\mathbf{I}),\mathbf{y}\right)
    +\lambda_c\|\Delta C\|_1\\
    &+\lambda_l\|\mathcal{M}_{31}(\Delta Y)\|_1
    +\lambda_r\|\mathbf{I}-\mathbf{b}\|_1
    +\lambda_a\mathcal{R}_{\mathrm{ctrl}},
\end{aligned}
\label{eq:local_objective}
\end{equation}
where $\Delta Y=\mathbf{w}^{\top}(\mathbf{I}-\mathbf{b})$ and $\Delta C=(\mathbf{I}-\mathbf{b})-\Delta Y\mathbf{1}$ denote luminance and chromatic changes, respectively. The control penalty is $\mathcal{R}_{\mathrm{ctrl}}=\frac{1}{3}\sum_k\operatorname{mean}(\mathbf{c}_k^2)$ over the three branches, where $\mathbf{c}_k$ contains the predicted coefficient maps of branch $k$. Here, $\|\cdot\|_1$ denotes the mean absolute value over tensor elements. We set $(\lambda_c,\lambda_l,\lambda_r,\lambda_a)=(0.1,0.1,0.01,0.001)$. These terms limit appearance changes and adjustment strengths, while task gradients guide local rendering on the frozen reconstruction frontend.

\section{Experiments}
\label{sec:experiments}

\begin{figure*}[t]
    \centering
    \includegraphics[width=\textwidth]{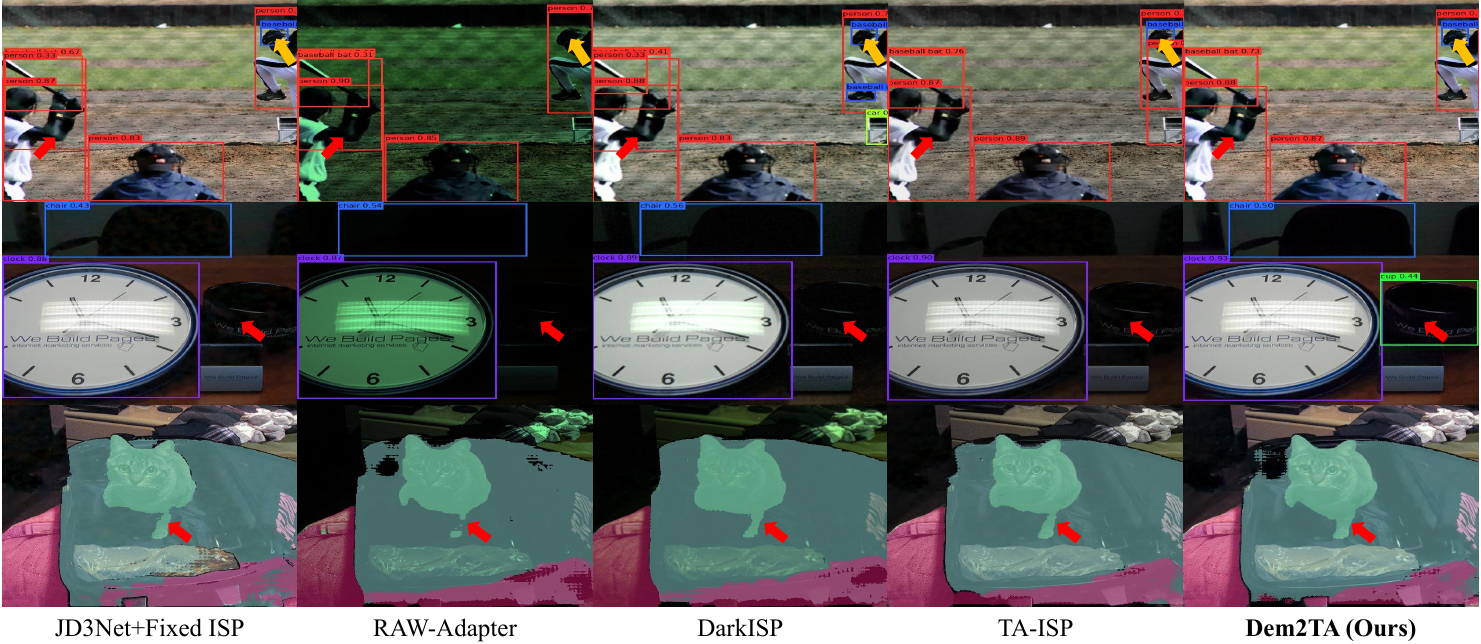}
    \caption{Qualitative detection and box-prompted segmentation comparisons on synthetic COCO RAW inputs. The examples illustrate the task-level benefits of jointly adapting reconstruction and perception, followed by local ISP optimization.}
    \label{fig:coco_results}
\end{figure*}

\begin{figure*}[t]
    \centering
    \includegraphics[width=\textwidth]{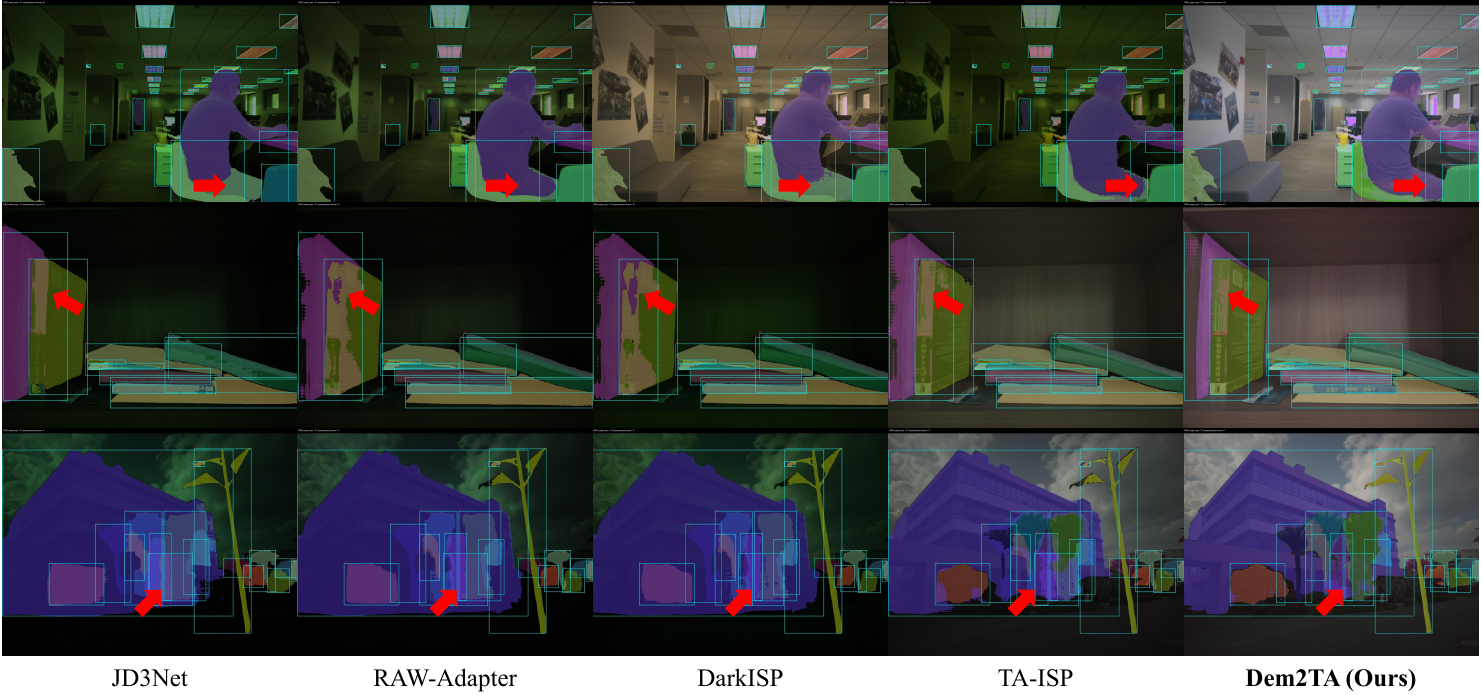}
    \caption{Instance segmentation on real RAW data. Quad Bayer images from MIPI 2022 and Single Bayer images captured with Huawei Mate 70, iPhone 14, and iPhone 18 Pro. We annotated all instances. Results on real data show that our method is generalizable.}
    \label{fig:results_real}
\end{figure*}

\begin{figure*}[t]
    \centering
    \includegraphics[width=\textwidth]{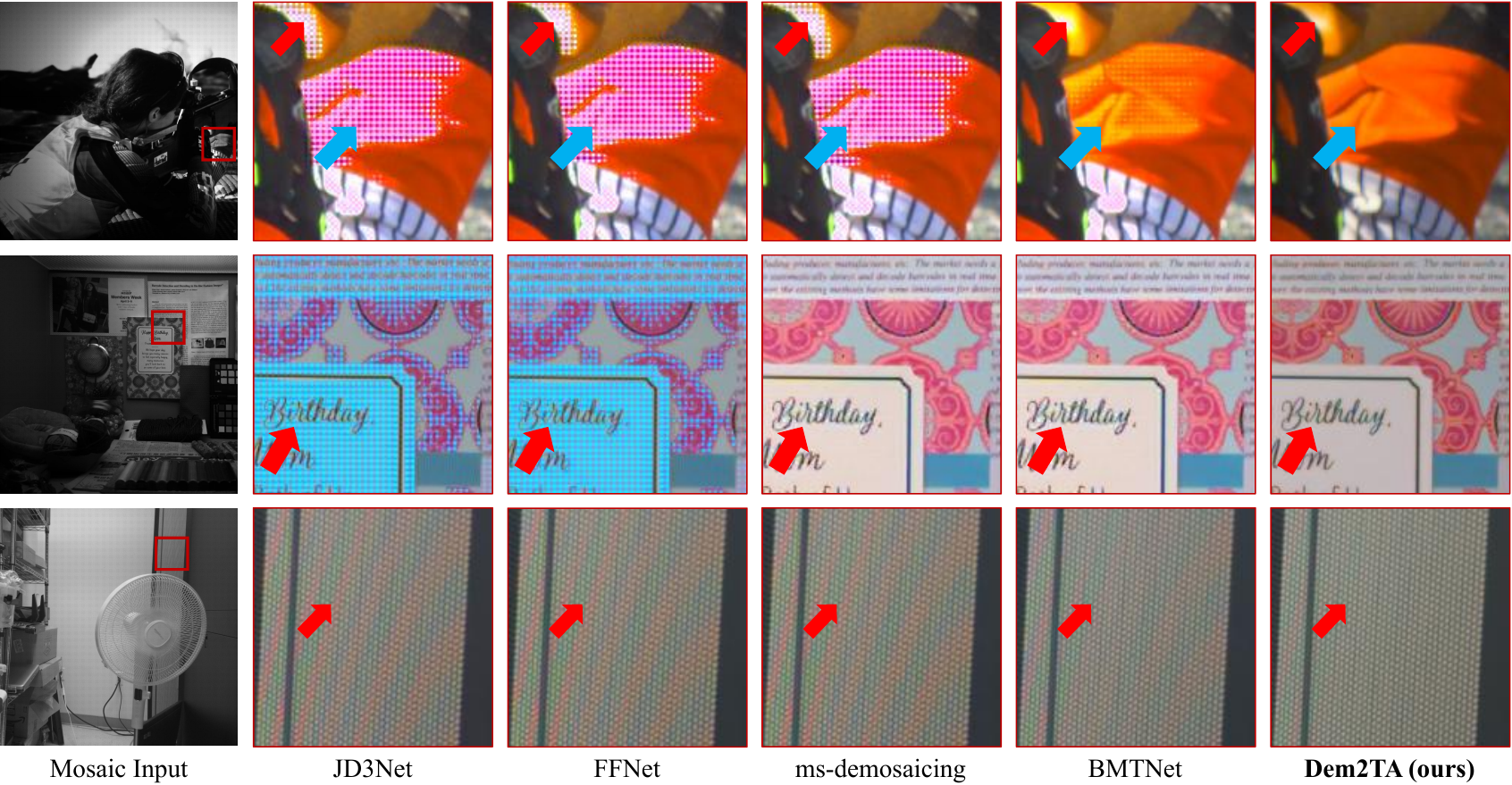}
    \caption{Demosaicing on real data using models trained on synthetic COCO. All outputs are rendered through the same fixed ISP.}
    \label{fig:demosaicing}
\end{figure*}

\subsection{Experimental Setup}
\label{sec:experiment_setup}

\paragraph{Tasks and evaluation.}
We evaluate object detection and box-prompted instance segmentation on COCO-derived~\cite{DBLP:conf/eccv/LinMBHPRDZ14} synthetic RAW images with RGGB Single-Bayer and RGGB Quad-Bayer layouts. RetinaNet~\cite{DBLP:conf/iccv/LinGGHD17} and MobileSAM~\cite{DBLP:journals/corr/abs-2306-14289} serve as the downstream models for detection and segmentation, respectively. The synthetic RAW images pipeline is followed by \cite{DBLP:journals/corr/abs-2608-07916}. Detection is evaluated using COCO AP, AP$_{50}$, AP$_{75}$, and AP for small, medium, and large objects. For box-prompted segmentation, we report mask IoU, Dice, precision, recall, Boundary F1, and Boundary IoU. Segmentation metrics evaluate prompted masks. Real-data collection and annotation are described in Sec.~\ref{sec:real_captured} and the supplementary material. Task adaptation uses the synthetic COCO training set; evaluation uses the synthetic test set and real-capture data. 

\paragraph{Baselines.}
We compare against RAW-Adapter~\cite{DBLP:conf/eccv/CuiH24}, Dark-ISP~\cite{DBLP:conf/iccv/GuoGYLP25}, AdaptiveISP~\cite{DBLP:conf/nips/WangXFXG24}, and TA-ISP~\cite{DBLP:conf/cvpr/ChenXZSG26}, using implementations adapted to our downstream protocols. In Table~\ref{tab:coco_comparison}, JD3Net~\cite{DBLP:conf/iccp/TedlaPZB25} and JD3Net + Fixed ISP are jointly fine-tuned with the downstream network to evaluate reconstruction-based pipelines; the fixed ISP applies white balance, color correction, and gamma mapping. RAW-Adapter and Dark-ISP use their native mosaic-input processing paths without an additional standalone demosaicing network. AdaptiveISP and TA-ISP are each evaluated with JD3Net and with the Stage-1-adapted Dem2TA (L) frontend. The Dem2TA-based variants and our local ISP start from the same Stage-1-trained demosaicing model and adapted downstream model. They use the same model architectures, experimental environment, subsequent training budget, and evaluation prompts, so this comparison evaluates the ISP strategies on a common adapted foundation. 

\subsection{Experiments on Synthetic Datasets}

\label{sec:synthetic_experiments}

In Table~\ref{tab:coco_comparison}, Dark-ISP exceeds JD3Net + TA-ISP in both tasks, whereas JD3Net + Fixed ISP achieves higher detection AP than Dark-ISP. Neither a dedicated demosaicing network nor a flexible ISP alone determines task performance. Using our adapted frontend and downstream initialization increases detection AP by 3.245 for AdaptiveISP and 5.259 for TA-ISP, although segmentation improves only for TA-ISP. The benefit of reconstruction therefore depends on its interaction with rendering and the downstream model.

From the same Stage-1-adapted foundation, our local ISP exceeds TA-ISP by 0.464 AP, 1.056 IoU, and 1.483 Boundary IoU, demonstrating the benefit of the second-stage rendering design. Dem2TA (S) slightly exceeds L in detection AP, while L performs better in segmentation and boundary metrics. These results support adapting reconstruction for perception while reconstruction accuracy directly do not predict task performance.

\begin{table}[htbp]
    \centering
    \caption{Segmentation results on the real-world dataset. The best and second-best results are shown in red and blue, respectively. $\dagger$ denotes using the Dem2TA (L) demosaicing model as its frontend.}
    \label{tab:real_segmentation}
    \setlength{\tabcolsep}{4pt}
    \renewcommand{\arraystretch}{1.1}

    \resizebox{\columnwidth}{!}{
    \begin{tabular}{llcccc}
        \toprule
        \multirow{2}{*}{Method}
        & \multirow{2}{*}{Reference}
        & \multicolumn{4}{c}{Segmentation} \\
        \cmidrule(lr){3-6}
        &
        & IoU$\uparrow$
        & Dice$\uparrow$
        & Boundary F1$\uparrow$
        & Boundary IoU$\uparrow$ \\
        \midrule

        RAW-Adapter
        & ECCV 2024
        & 82.468 & 89.629 & 50.545 & 14.508 \\

        AdaptiveISP$^{\dagger}$
        & NeurIPS 2024
        & 79.636 & 88.512 & 34.971 & 10.908 \\

        Dark-ISP
        & ICCV 2025
        & 82.331 & 89.490 & 50.850 & 15.012 \\

        TA-ISP$^{\dagger}$
        & CVPR 2026
        & 82.890 & \bluenum{89.838} & 51.934 & 14.640 \\

        \midrule

        Dem2TA (L)
        & Ours
        & \rednum{83.519} & \rednum{90.236}
        & \rednum{53.964} & \rednum{15.773} \\

        Dem2TA (S)
        & Ours
        & \bluenum{82.939} & 89.831
        & \bluenum{53.108} & \bluenum{15.417} \\

        \bottomrule
    \end{tabular}
    }
\end{table}

\subsection{Experiments on Real Captured Datasets}
\label{sec:real_captured}

We evaluate models trained on synthetic RAW data using 100 real images (63 Quad Bayer and 37 Single Bayer) with 1,112 annotated instances. The set combines MIPI 2022~\cite{DBLP:conf/eccv/YangYJLFZSZLGWLZPCZWWFXZZFZLATPLJSZXXJ22} with Huawei Mate 70, iPhone 14, and iPhone 18 Pro captures~\cite{huawei_mate70,apple_iphone14,apple_iphone18pro}. Candidate masks generated by SAM 3~\cite{DBLP:conf/iclr/CarionGHDHCRAKH26} are manually reviewed and refined in CVAT. Each target has a separate instance mask, and all instances share the category \texttt{object}. Collection, annotation, and quality-control details are provided in the supplementary material.

Table~\ref{tab:real_segmentation} shows that Dem2TA (L) leads all four metrics. Compared with TA-ISP using the same frontend, it improves IoU by 0.629 and Boundary F1 by 2.030, extending the benefits of local adaptation to real captures. S ranks second in IoU and both boundary metrics. The gains extend beyond synthetic data, as also illustrated in Fig.~\ref{fig:results_real}.

\subsection{Ablation Study}
\label{sec:ablation}

\paragraph{Comparison on Demosaicing Baselines.}

We compare SANet~\cite{zhang2022sanet}, JD3Net~\cite{DBLP:conf/iccp/TedlaPZB25}, BMTNet~\cite{DBLP:conf/cvpr/ZhouZLSTCLS25}, FFNet~\cite{math14071175}, and ms-demosaic~\cite{DBLP:conf/iccv/TedlaLYAB25} on synthetic COCO data with both CFA layouts. Table~\ref{tab:demosaicing_reconstruction} also compares Dem2TA before and after Stage~1: the former uses original MobileSAM, while the latter uses the jointly adapted frontend and downstream model. Both use the same fixed ISP without local enhancement. Qualitative results on MIPI 2022 is shown in Fig.~\ref{fig:demosaicing}.

The reconstruction baselines have substantially different PSNR values but closely grouped segmentation scores. Joint frontend--downstream adaptation improves IoU by 3.099 for L and 3.431 for S, with Boundary F1 gains of 1.530 and 3.463. L retains its reconstruction quality, whereas S gains task accuracy despite lower reconstruction scores. These results support adapting reconstruction and perception as a pipeline, rather than treating reconstruction quality as a direct predictor of task accuracy. Further analysis are provided in the supplementary material.

\begin{table}[htbp]
    \centering
    \caption{Reconstruction and segmentation results before and after joint adaptation, averaged over dual Bayer inputs. Segmentation uses original MobileSAM before adaptation and the jointly adapted model afterward, with a fixed ISP throughout. Best and second-best results are highlighted in red and blue, respectively.}
    \label{tab:demosaicing_reconstruction}
    \setlength{\tabcolsep}{2.5pt}
    \renewcommand{\arraystretch}{1.12}
    \resizebox{\columnwidth}{!}{%
    \begin{tabular}{llr*{7}{r}}
        \toprule
        \multirow{2}{*}{Method}
        & \multirow{2}{*}{Reference}
        & \multirow{2}{*}{Param. (M)}
        & \multicolumn{2}{c}{Linear RGB}
        & \multicolumn{3}{c}{sRGB}
        & \multicolumn{2}{c}{Segmentation} \\
        \cmidrule(lr){4-5}
        \cmidrule(lr){6-8}
        \cmidrule(lr){9-10}
        & &
        & PSNR$\uparrow$ & SSIM$\uparrow$
        & PSNR$\uparrow$ & SSIM$\uparrow$ & LPIPS$\downarrow$
        & IoU$\uparrow$ & BF1$\uparrow$ \\
        \midrule
        SANet & AAAI 2022 & 5.72
        & 40.1412 & 0.9719 & 30.2012 & 0.8554 & 0.2006
        & 76.889 & 66.485 \\

        BMTNet & CVPR 2025 & 12.17
        & 42.2246 & 0.9581 & 31.4946 & 0.8689 & 0.1779
        & 76.921 & 66.562 \\

        ms-demosaic & ICCV 2025 & 10.68
        & 43.2656 & 0.9798 & 33.7102 & 0.9025 & 0.1265
        & 76.917 & 66.541 \\

        JD3Net & ICCP 2025 & 3.20
        & 41.2850 & 0.9838 & 31.7322 & 0.9146 & 0.1302
        & 76.913 & 66.580 \\

        FFNet & MDPI 2026 & 43.26
        & 41.6340 & 0.9744 & 31.5293 & 0.8756 & 0.1778
        & 76.907 & 66.502 \\

        \midrule
        \rowcolor{gray!15}
        \multicolumn{10}{l}{\textit{Before segmentation adaptation}} \\

        Dem2TA (L) & Ours & 33.69
        & \bluenum{44.6938}
        & \bluenum{0.9882}
        & \bluenum{34.7775}
        & \bluenum{0.9221}
        & \rednum{\textbf{0.1045}}
        & 76.924 & 66.591 \\

        Dem2TA (S) & Ours & 6.65
        & 44.2218 & 0.9879 & 34.1961 & 0.9160 & 0.1118
        & 77.064 & 66.518 \\

        \midrule
        \rowcolor{gray!15}
        \multicolumn{10}{l}{\textit{After segmentation adaptation}} \\

        Dem2TA (L) & Ours & 33.69
        & \rednum{\textbf{44.7186}}
        & \rednum{\textbf{0.9883}}
        & \rednum{\textbf{34.8519}}
        & \rednum{\textbf{0.9244}}
        & \bluenum{0.1059}
        & \bluenum{80.023}
        & \bluenum{68.121} \\

        Dem2TA (S) & Ours & 6.65
        & 42.7870 & 0.9829 & 31.0417 & 0.8571 & 0.1723
        & \rednum{\textbf{80.495}}
        & \rednum{\textbf{69.981}} \\

        \bottomrule
    \end{tabular}%
    }
\end{table}

\paragraph{Ablation Study on Demosaicing Adaptation}

Table~\ref{tab:ablation_demo_seg} examines adaptation settings. Downstream-only fine-tuning freezes demosaicing; joint fine-tuning updates both models using reconstruction and task losses without representation alignment. We then add DINO-only, task-only, or both alignment signals with the same fixed ISP. The frozen alignment encoder is a separate copy, while the tuned encoder shares weights with the downstream encoder. Full adds sequential local ISP adaptation; Joint stages updates the frontend and local ISP simultaneously.

Simply updating the frontend alongside the downstream model yields a limited IoU gain of 0.117 over downstream-only tuning. Representation-guided joint adaptation improves IoU by a further 2.644, showing how representation supervision helps the reconstruction stage participate effectively in pipeline adaptation. DINO-only and task-only alignment gain 0.482 and 1.653 over ordinary joint fine-tuning; combining them exceeds task-only alignment by 0.991 IoU, with a 0.343 Boundary F1 trade-off. Sequential training slightly exceeds simultaneous updates by 0.105 IoU and 0.370 Boundary F1, supporting this training arrangement in the evaluated setting.

\begin{table}[t]
    \centering
    \caption{Ablation of adaptation components on COCO instance segmentation. ``Encoder frozen'' uses a separate frozen alignment copy; ``encoder tuned'' shares weights with the trainable downstream encoder. ``Full'' adds sequential local ISP adaptation; ``Joint stages'' updates the demosaicing frontend and local ISP together.}
    \label{tab:ablation_demo_seg}
    \setlength{\tabcolsep}{3.2pt}
    \renewcommand{\arraystretch}{1.12}
    \resizebox{\columnwidth}{!}{%
    \begin{tabular}{lccccc}
        \toprule
        Setting & IoU$\uparrow$ & Dice$\uparrow$
        & Prec.$\uparrow$ & BF1$\uparrow$ & BIoU$\uparrow$ \\
        \midrule
        Pretrained                    & 76.924 & 85.512 & 87.498 & 66.591 & 26.201 \\
        Downstream-only fine-tune                     & 77.262 & 85.691 & 87.730 & 66.839 & 26.305 \\
        Joint fine-tune, no alignment & 77.379 & 85.780 & 87.894 & 67.011 & 26.412 \\
        + DINO only                   & 77.861 & 85.880 & 88.130 & 67.122 & 26.451 \\
        + Task only, encoder tuned    & 79.032 & 87.001 & 89.231 & 68.464 & 26.511 \\
        + DINO + Task, encoder frozen & 79.593 & 88.045 & 90.699 & 68.914 & 26.162 \\
        + DINO + Task, encoder tuned  & 80.023 & 87.508 & \textbf{91.483} & 68.121 & 26.879 \\
        \midrule
        Joint stages                  & 81.047 & 88.577 & 90.099 & 71.200 & 27.985 \\
        \textbf{Full}                 & \textbf{81.152} & \textbf{88.640}
                                      & 90.114 & \textbf{71.570} & \textbf{28.153} \\
        \bottomrule
    \end{tabular}%
    }
\end{table}

\paragraph{Ablation Study on Local ISP}

\begin{figure}[htbp]
    \centering
    \includegraphics[width=\linewidth]{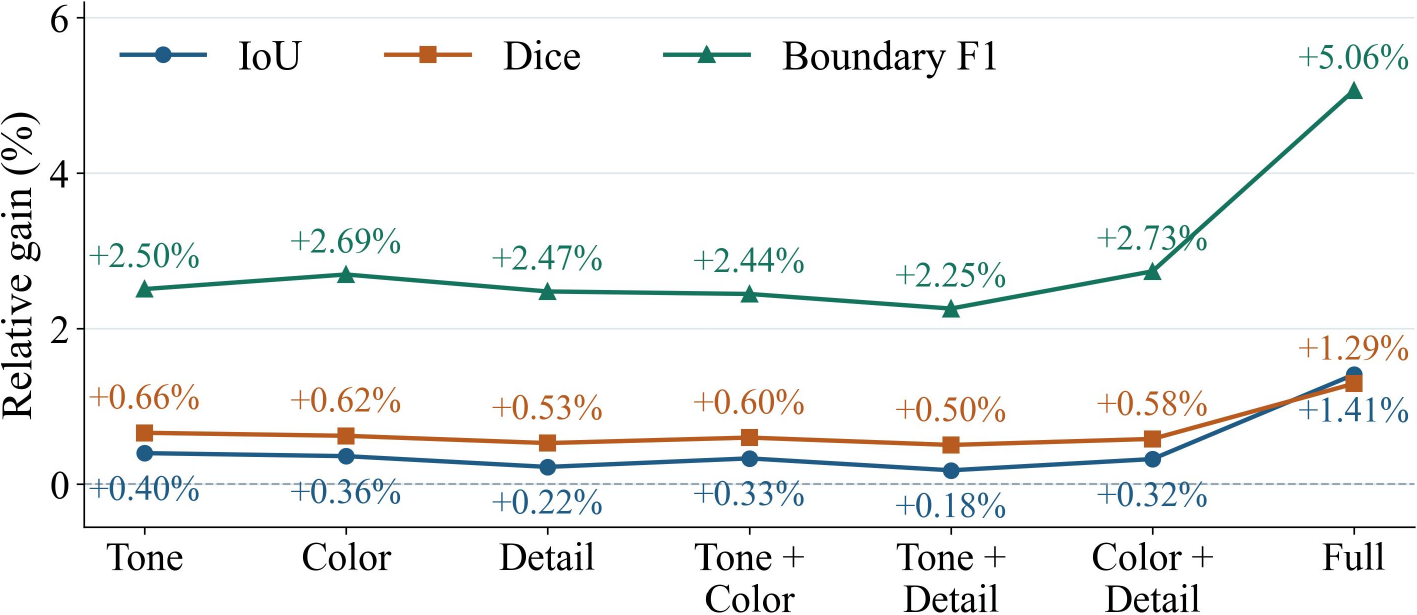}
    \caption{Relative gains over the fixed-ISP reference on COCO box-prompted segmentation. Individual controls provide gains, while the full local ISP achieves the largest improvements across all three metrics, supporting their joint use for prediction.}
    \label{fig:ISP_ablation}
\end{figure}

Starting from the same converged Stage-1 checkpoint, we freeze the demosaicing frontend in all variants. The fixed-ISP reference continues training only the downstream model, while the local-ISP variants jointly optimize the rendering controls and downstream model. All variants use the same training data and number of additional training epochs. Figure~\ref{fig:ISP_ablation} reports relative gains.

Each control improves segmentation when used alone. Tone gives the largest individual IoU gain of 0.319, while color gives the largest Boundary F1 gain of 1.834. Pairwise gains are not additive: tone and color use luminance-dependent modulation, while detail bands are extracted from the same luminance, so their adjustments can affect overlapping regions. Joint optimization of the full local ISP and downstream model improves IoU, Dice, and Boundary F1 by 1.129, 1.132, and 3.449, respectively. These results support further improvements through local ISP adaptation.

\section{Conclusion}

We presented Dem2TA, a framework that incorporates explicit, fidelity supervised demosaicing into task-oriented ISP adaptation. Complementary semantic and task representation alignment guides the joint adaptation of the reconstruction frontend and downstream model, improving perception beyond joint fine-tuning alone. A lightweight local ISP subsequently refines task-specific rendering while keeping the adapted reconstruction fixed. Experiments on synthetic RAW detection and box-prompted segmentation, together with real-capture segmentation evaluation, support the effectiveness of this framework. Our results also highlight that pixel fidelity alone is insufficient to predict downstream performance, and that reconstruction–perception trade-offs depend on the frontend configuration.

\section{LLM Usage Statement}

We used OpenAI's ChatGPT/GPT tools to assist with manuscript drafting and editing, literature exploration, code development, experimental analysis, and figure preparation, and to obtain reviewer-style feedback on the manuscript. All AI-assisted content was reviewed and verified by the authors, who take full responsibility for the final manuscript, implementation, experimental results, and conclusions.

{
    \small
    \bibliographystyle{ieeenat_fullname}
    \bibliography{main}
}

\clearpage
\setcounter{page}{1}
\maketitlesupplementary

\section{Real Dataset Collection and Labeling}
\label{sec:supp_real_dataset}

\begin{figure}[htbp]
    \centering
    \includegraphics[width=0.49\textwidth]{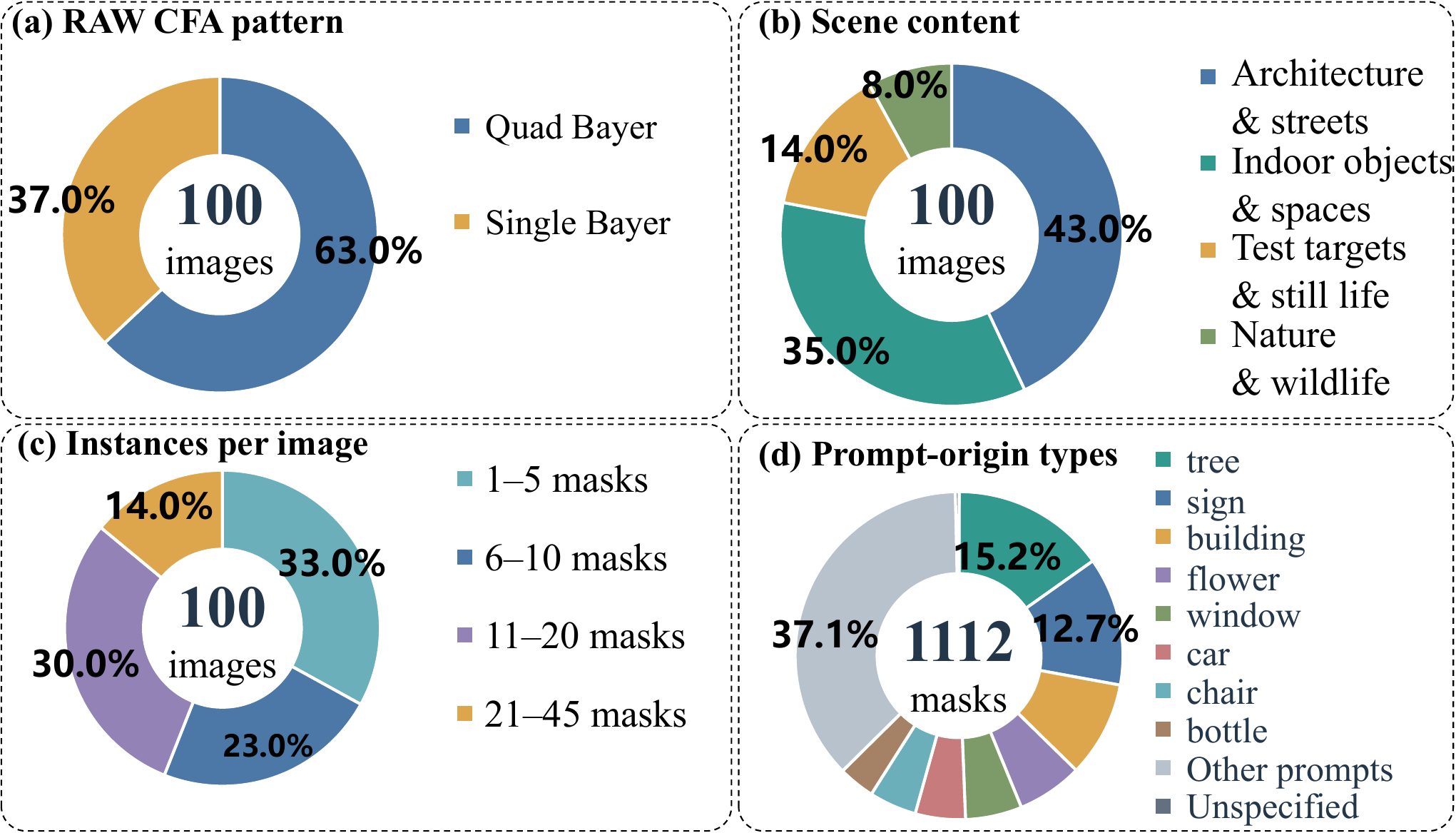}
    \caption{Composition of the real-capture evaluation set: (a) RAW CFA patterns, (b) scene content, (c) numbers of instances per image, and (d) types of prompts used to generate candidate masks.}
    \label{fig:supp_dataset_composition}
\end{figure}

\paragraph{Dataset composition.}
Our real-capture evaluation set contains 100 images with 1,112 annotated instances. As shown in Fig.~\ref{fig:supp_dataset_composition}(a), 63 images use the Quad Bayer pattern and 37 use the Single Bayer pattern. The images cover a range of scenes, including architecture and streets, indoor objects and spaces, test targets and still life, and nature and wildlife. Their proportions are 43\%, 35\%, 14\%, and 8\%, respectively (Fig.~\ref{fig:supp_dataset_composition}(b)). The number of annotated instances also varies across images: 33 images contain 1--5 instances, 23 contain 6--10, 30 contain 11--20, and 14 contain 21--45 (Fig.~\ref{fig:supp_dataset_composition}(c)).

\begin{figure*}[t]
    \centering
    \includegraphics[width=\textwidth]{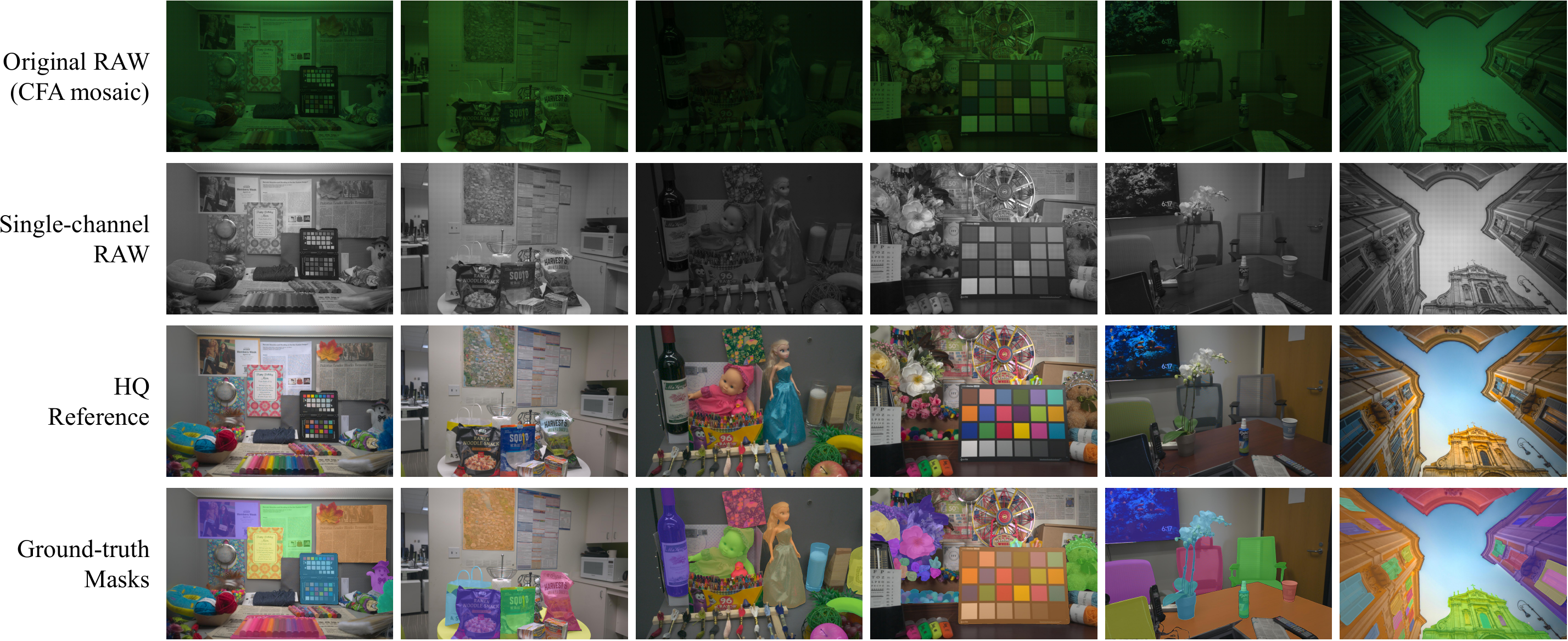}
    \caption{Examples of Quad Bayer data and instance annotations. From top to bottom: CFA mosaic visualization, single-channel RAW, RGB reference, and ground-truth instance masks. Different mask colors distinguish instances.}
    \label{fig:supp_quad_bayer}
\end{figure*}

\begin{figure*}[t]
    \centering
    \includegraphics[width=\textwidth]{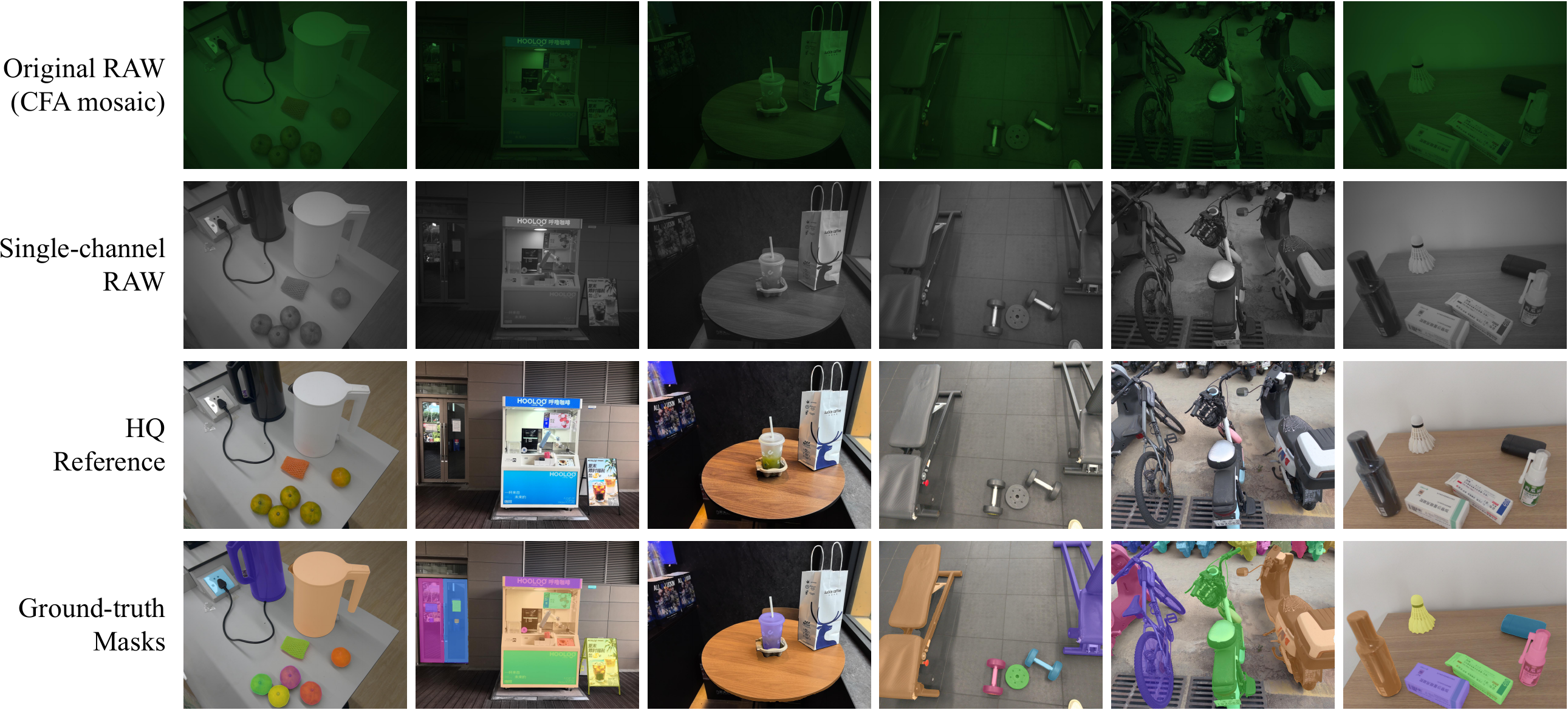}
    \caption{Examples of Single Bayer data and instance annotations. From top to bottom: CFA mosaic visualization, single-channel RAW, RGB reference, and ground-truth instance masks. Different mask colors distinguish instances.}
    \label{fig:supp_single_bayer}
\end{figure*}

\paragraph{Data collection and reference images.}
The dataset comprises 63 Quad Bayer images from MIPI 2022~\cite{DBLP:conf/eccv/YangYJLFZSZLGWLZPCZWWFXZZFZLATPLJSZXXJ22} and two Single Bayer mobile-phone subsets
containing 30 and 7 images. All 100 images have corresponding RAW data, RGB references, and instance annotations. The MIPI references are rendered from paired Bayer
ground truth, whereas the phone references are the corresponding camera JPEGs. Annotation is performed on these RGB references rather than on RAW visualizations.
Representative examples are shown in Figs.~\ref{fig:supp_quad_bayer} and~\ref{fig:supp_single_bayer}.

\paragraph{RAW representation and preprocessing.}
All inputs retain a single-channel CFA mosaic; the phone DNGs are not demosaiced Linear DNGs. RAW values are normalized as
$u=\operatorname{clip}((\mathrm{DN}-b)/(w-b),0,1)$.
The black/white levels $(b,w)$ are $(64,1023)$ for MIPI and either $(528,4095)$ or $(256,4095)$ for the phone files, according to their metadata. MIPI inputs
additionally use green-median exposure normalization toward 0.135, with the scale clipped to $[0.05,8]$. Single Bayer follows RGGB sampling, while Quad Bayer arranges
$2\times2$ same-color blocks in RGGB order. Pattern-aware packing and same-color binning provide a common four-channel RAW condition for the restoration network.

\paragraph{RAW--RGB spatial correspondence.}
MIPI inputs and rendered references share the original pixel grid. For each phone pair, an affine transformation is estimated using SIFT and RANSAC between a rendered
DNG preview and the reference JPEG, without using annotation masks. The RAW mosaic is not globally resized or rotated; only bottom/right reflection padding is applied
where required. After restoration, padding is removed and RGB is mapped to the reference-image coordinates. During segmentation, RGB and box prompts are consistently
scaled to the model input, and predicted masks are mapped back to the original reference grid for evaluation.

\begin{figure*}[t]
    \centering
    \includegraphics[width=\textwidth]{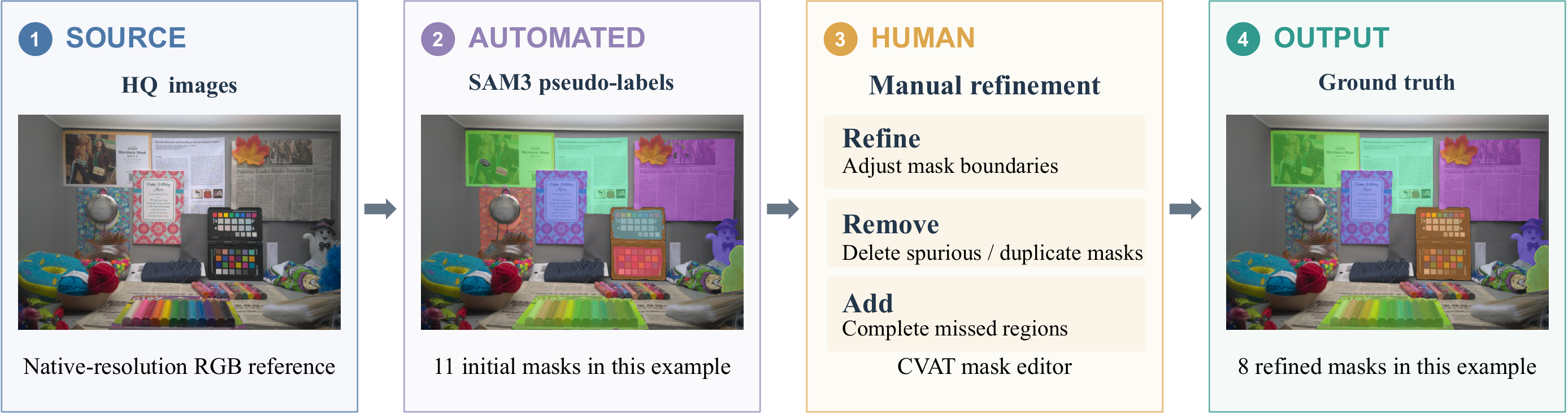}
    \caption{Semi-automatic annotation workflow. SAM 3 generates candidate masks on an RGB reference image, which are then refined in CVAT to produce instance-level ground truth.}
    \label{fig:supp_annotation_workflow}
\end{figure*}

\paragraph{Annotation protocol.}
We create pixel-level instance masks on the RGB reference images using SAM 3 for pre-annotation followed by manual refinement in CVAT, as illustrated in Fig.~\ref{fig:supp_annotation_workflow}. The annotations are class-agnostic: each target is represented by a separate instance mask, and all instances share the category \texttt{object}. Text prompts used during pre-annotation describe candidate targets but are not treated as verified semantic labels. Figure~\ref{fig:supp_dataset_composition}(d) summarizes the prompt-origin types associated with the masks; these types should therefore not be interpreted as semantic ground-truth categories.

\paragraph{Automatic pre-annotation.}
SAM 3 is applied to each RGB reference image with text prompts describing the targets to be annotated. Candidate masks are filtered by confidence, size, and redundancy, and are stored with image metadata and instance identifiers in COCO format. These masks serve as an initial proposal set and are manually checked before inclusion in the ground truth. The example in Fig.~\ref{fig:supp_annotation_workflow} contains 11 initial candidate masks and eight masks after refinement; these numbers describe that example rather than the full dataset.

\paragraph{Manual refinement.}
Annotators inspect the original images and candidate masks in CVAT. Mask boundaries are refined using brush and eraser tools, erroneous or duplicate masks are removed, and missed targets within the annotation scope are added. Distinguishable objects are kept as separate instances. Boundaries follow visible object contours, and genuinely occluded regions are not inferred. The original image content, filenames, and spatial resolutions are preserved, together with the verified RAW--RGB pairing
for all 100 images.

\paragraph{Export and quality control.}
The reviewed annotations are exported in COCO format. We check image filenames and dimensions, image--annotation associations, instance identifiers, mask decoding, and mask areas. For masks representing individual instances, the \texttt{iscrowd} field is set to zero when necessary; genuine crowd annotations are not changed indiscriminately. Finally, exported masks are overlaid on the reference images to check spatial alignment, missing or duplicate instances, and boundary errors. Images confirmed to contain no targets are retained without artificial annotations. Only annotations that pass manual review and final checks are used for evaluation.

\section{Dem2TA Demosaicing Adaptation}
\label{sec:supp_adaptation}

\subsection{Multi-Scale Fusion Implementation}
\label{sec:supp_fusion_details}

We aggregate features from all four demosaicing decoder levels using $1\times1$ channel projections and spatial downsampling. The two finest-resolution features are concatenated and projected into a shared level, forming a three-level pyramid. Self-attention is applied at the coarsest scale, followed by top-down and bottom-up fusion~\cite{DBLP:conf/cvpr/ZhaoLXWWDLC24}. The top-down path upsamples coarse features and concatenates them with finer features, while the bottom-up path uses stride-2 convolutions to propagate information back to coarser resolutions.

Each fusion node employs a cross-stage partial reparameterizable block (CSP-Rep)~\cite{DBLP:conf/cvpr/ZhaoLXWWDLC24}. Two parallel $1\times1$ projections form a transformation branch and a shortcut branch. The transformation branch contains Rep blocks, each combining $3\times3$ and $1\times1$ convolutional branches. The transformed features are added to the shortcut features and projected to the output dimension. Finally, two separate projection heads map the fused pyramid into the DINO and task-encoder feature spaces. The fusion branches provide training-time supervision and are not used to produce task predictions at inference.

\paragraph{Reference encoders and alignment.}
\label{sec:supp_alignment_settings}
Reference features are extracted from sRGB images. DINOv2 remains frozen and supplies channel-normalized tokens for Smooth-$L_1$ alignment. Task references come from the MobileSAM image encoder for segmentation and the RetinaNet feature extractor for detection; the corresponding projection head is supervised using mean squared error. Separate heads accommodate the two feature spaces without directly aligning DINO and task embeddings to each other. In the ``encoder frozen'' ablation, the alignment target is produced by a separately copied, frozen task encoder; this copy receives no alignment or task gradients, while the downstream network is still fine-tuned. In the ``encoder tuned'' setting, the alignment task encoder and the corresponding downstream encoder share weights and are updated together, including through the task-alignment loss. Thus, the task target can evolve during training in this setting; ``frozen'' refers only to the separate alignment copy, not to the downstream network. The downstream network processes fixed-ISP renderings of reconstructed linear RGB. Both alignment branches are used only during training.

\paragraph{Lightweight frontend.}
Dem2TA (L) uses the Restormer-based reconstruction frontend. Dem2TA (S) uses a lightweight convolutional variant built with depthwise convolutions. Both follow the same two-stage adaptation framework; their reconstruction and downstream results are reported separately.

\subsection{Why Adapt Demosaicing for Perception?}
\label{sec:discussion_adaptation}

Our goal is to make reconstruction more useful for perception, rather than assume that higher reconstruction scores necessarily lead to better task performance. A pretrained demosaicing model provides a fidelity prior, but its features are primarily learned to recover missing color samples. Our adaptation introduces semantic and task-related guidance into these features while jointly updating the downstream network.

\paragraph{Reconstruction quality is not a sufficient predictor of perception.}
Table~\ref{tab:coco_comparison} shows that the benefit of a reconstruction frontend depends on the subsequent ISP and task. Starting from our Stage-1-adapted frontend and downstream model improves detection AP by 3.245 points for AdaptiveISP and 5.259 points for TA-ISP over their JD3Net-based counterparts. However, segmentation IoU increases by 1.008 points for TA-ISP but decreases by 1.865 points for AdaptiveISP. Since these comparisons also involve different downstream initializations, they evaluate the adapted foundation rather than isolate reconstruction quality alone. Moreover, JD3Net with the fixed ISP achieves higher detection AP than our frontend with AdaptiveISP. These results motivate adapting the relationship between reconstruction and perception, rather than relying on frontend quality alone.

This distinction can be expressed through task sensitivity. Let $\hat{\mathbf{z}}=D_{\theta}(\mathbf{x})$ denote reconstructed linear RGB and let
\begin{equation}
    \mathcal{L}_{\mathrm{task}}(\hat{\mathbf{z}})
    =
    \ell_{\mathrm{task}}
    \bigl(T_{\psi}(\mathcal{P}(\hat{\mathbf{z}})),\mathbf{y}\bigr).
\end{equation}
For a small perturbation $\boldsymbol{\delta}$ to the reconstruction, with the ISP and task-network parameters held fixed,
\begin{equation}
    \mathcal{L}_{\mathrm{task}}(\hat{\mathbf{z}}+\boldsymbol{\delta})
    -
    \mathcal{L}_{\mathrm{task}}(\hat{\mathbf{z}})
    \approx
    \left\langle
    \nabla_{\hat{\mathbf{z}}}\mathcal{L}_{\mathrm{task}},
    \boldsymbol{\delta}
    \right\rangle.
    \label{eq:task_sensitive_reconstruction}
\end{equation}
Pixel reconstruction metrics measure discrepancies from the reference, whereas task loss also depends on where and in which direction image changes occur. Their improvements therefore need not be monotonic. This observation motivates retaining reconstruction supervision while explicitly learning task-relevant representations.

\paragraph{Complementary alignment connects reconstruction and perception.}
Let $\mathbf{F}_{\theta}(\mathbf{x})$ denote the multiscale demosaicing features. We align their projections with DINO and task-encoder features extracted from reference sRGB images $\mathbf{s}$:
\begin{equation}
\begin{aligned}
    \mathcal{L}_{\mathrm{align}}
    ={}&
    \lambda_d\,d_d
    \bigl(A_d(\mathbf{F}_{\theta}(\mathbf{x})),E_d(\mathbf{s})\bigr)\\
    &+
    \lambda_e\,d_e
    \bigl(A_e(\mathbf{F}_{\theta}(\mathbf{x})),E_e(\mathbf{s})\bigr).
\end{aligned}
\end{equation}
Here, $A_d$ and $A_e$ accommodate the different reference spaces. Frozen DINO provides a shared semantic reference, while task-encoder features supply task-related spatial guidance. The alignment acts through the reconstruction hierarchy; it does not require DINO and task embeddings to match each other directly.

Table~\ref{tab:ablation_demo_seg} supports the complementary role of these signals. Relative to downstream-only fine-tuning with demosaicing frozen, joint adaptation with DINO-only or task-only alignment improves IoU by 0.599 or 1.770 points. These contrasts include both frontend updates and representation supervision, so they evaluate the corresponding adaptation strategies together. Combining both with a fine-tuned alignment task encoder reaches 80.023 IoU, exceeding task-only alignment by 0.991 points. With both references present, fine-tuning the alignment task encoder instead of freezing it improves IoU by 0.430 points, while Dice and Boundary F1 decrease by 0.537 and 0.793 points. The combined references give the strongest Stage-1 IoU when the task encoder adapts, but the other metrics show a trade-off. Subsequent local ISP optimization improves all three metrics over this adapted fixed-ISP endpoint.

The frontend receives reconstruction, alignment, and task gradients:
\begin{equation}
    \nabla_{\theta}\mathcal{L}_{\mathrm{adapt}}
    =
    \lambda_{\mathrm{rec}}\nabla_{\theta}\mathcal{L}_{\mathrm{rec}}
    +
    \nabla_{\theta}\mathcal{L}_{\mathrm{align}}
    +
    \lambda_{\mathrm{task}}\nabla_{\theta}\mathcal{L}_{\mathrm{task}}.
\end{equation}
These signals constrain different aspects of learning: reconstruction supervision preserves image information, alignment organizes internal features, and task supervision connects them to downstream predictions. Meanwhile, the downstream network learns to process the adapted images. This joint learning process is the basis of our reconstruction--perception interaction.

\paragraph{Fidelity preservation depends on the model configuration.}
Table~\ref{tab:demosaicing_reconstruction} measures reconstruction and downstream performance together. For L, segmentation adaptation improves linear-RGB PSNR by 0.0248~dB and sRGB PSNR by 0.0744~dB, while IoU and Boundary F1 increase by 3.099 and 1.530 points. LPIPS changes from 0.1045 to 0.1059, so fidelity is broadly retained rather than improved in every metric. For S, linear-RGB and sRGB PSNR decrease by 1.4348~dB and 3.1544~dB, yet IoU and Boundary F1 increase by 3.431 and 3.463 points. These within-model comparisons use the fixed ISP and isolate the outcome of joint frontend--downstream adaptation from subsequent learnable rendering. They show why the pipeline must be evaluated with both reconstruction and task metrics: improved perception need not coincide with improved pixel fidelity.

The lightweight convolutional architecture has less capacity than the larger Restormer frontend, which retains transposed channel attention and gated feed-forward blocks. This may contribute to a stronger trade-off between reconstruction and task adaptation, but the comparison does not isolate architecture as the cause. The results show that fidelity preservation is not uniform across model configurations and should be measured alongside downstream performance.

\paragraph{Frozen downstream diagnostic.}
We evaluate the pretrained and Stage-1-adapted demosaicing frontends with the same original, frozen MobileSAM and fixed ISP. The COCO validation protocol uses 2,342 pattern-image samples and one deterministic ground-truth box per sample. No local ISP or adapted MobileSAM weights are loaded. Substituting the Stage-1 frontend leaves segmentation performance nearly unchanged for both L and S. In contrast, the jointly adapted pipeline shows substantial task gains in Table~\ref{tab:demosaicing_reconstruction}; this comparison does not attribute those gains to the frontend alone.

\begin{table}[htbp]
    \centering
    \caption{Frontend-only substitution with the original MobileSAM frozen on COCO box-prompted segmentation. Both rows for each model share the fixed ISP, prompts, and evaluation protocol.}
    \label{tab:supp_frozen_sam}
    \setlength{\tabcolsep}{4pt}
    \begin{tabular}{llrrr}
        \toprule
        Frontend & Stage 1 & IoU$\uparrow$ & Dice$\uparrow$ & Boundary F1$\uparrow$ \\
        \midrule
        L & Before & 77.182 & 85.635 & 66.778 \\
        L & After  & 77.218 & 85.655 & 66.818 \\
        S & Before & 77.064 & 85.545 & 66.518 \\
        S & After  & 76.997 & 85.501 & 66.361 \\
        \bottomrule
    \end{tabular}
\end{table}

\paragraph{Local ISP adaptation builds on the adapted reconstruction.}
Stage 2 freezes the demosaicing frontend and optimizes local rendering with the downstream network. In Table~\ref{tab:ablation_demo_seg}, the complete sequential framework improves over the Stage-1 encoder-tuned setting by 1.129 IoU points and 3.449 Boundary F1 points. Compared with simultaneous frontend and local-ISP updates, sequential adaptation achieves an additional 0.105 IoU and 0.370 Boundary F1 points. These modest gains favor the sequential schedule in the evaluated setting.

The same-foundation comparison in Table~\ref{tab:coco_comparison} further shows that our local ISP exceeds TA-ISP by 0.464 detection AP, 1.056 segmentation IoU, and 1.483 Boundary IoU points. Together, the results support the two complementary roles of our framework: Stage 1 adapts reconstruction representations and their use by the downstream model, while Stage 2 refines task-oriented rendering without further changing the reconstructed linear RGB.

\subsection{Local ISP Parameter Efficiency}
\label{sec:supp_isp_parameters}

\begin{figure}[htbp]
    \centering
    \includegraphics[width=\linewidth]{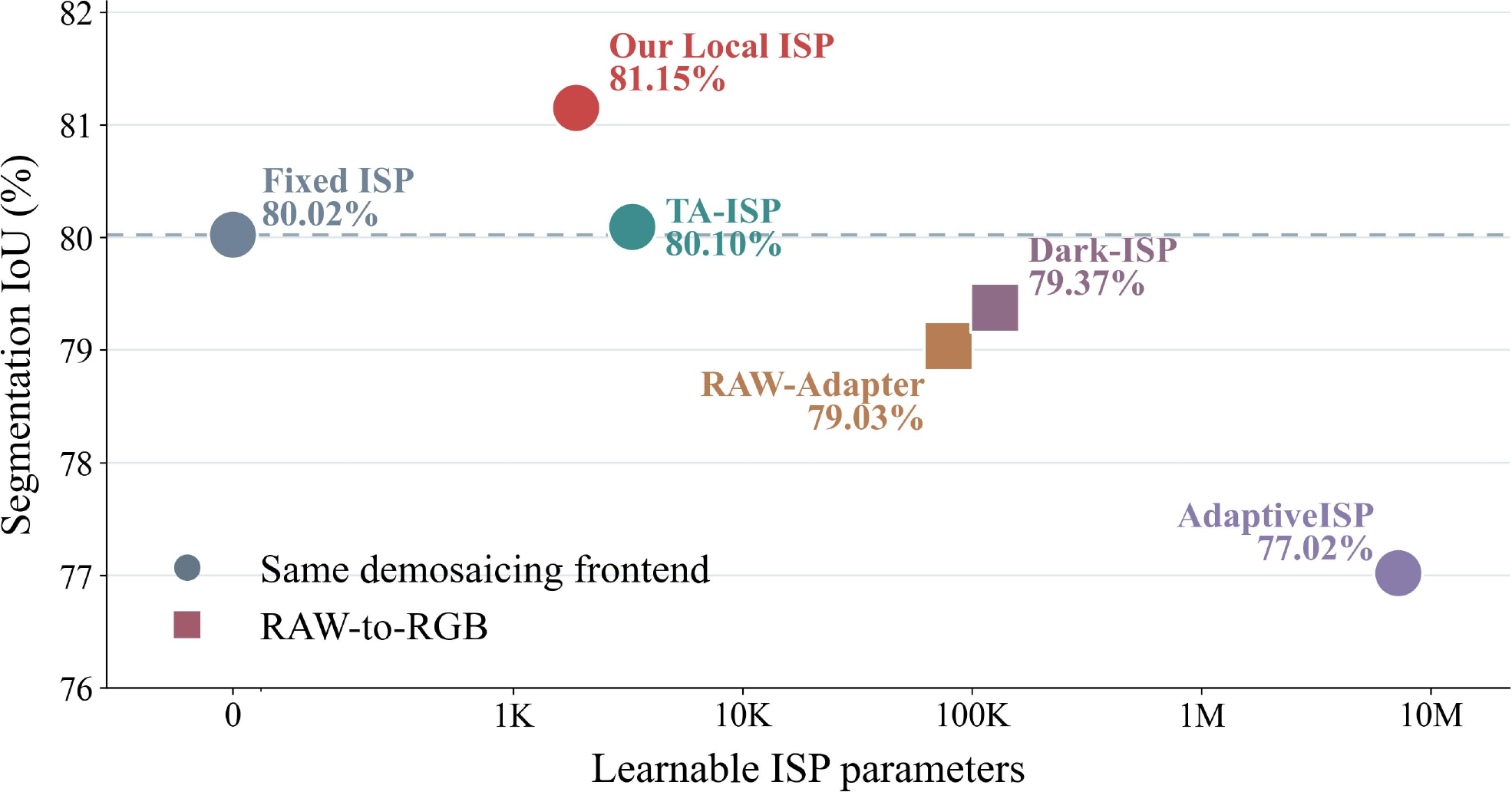}
    \caption{Our local ISP achieves competitive performance with a small number of parameters. Circles denote methods using the same demosaicing frontend, while squares denote RAW-to-RGB methods. The dashed line indicates the performance of Fixed ISP.}
    \label{fig:ISP}
\end{figure}

Figure~\ref{fig:ISP} compares segmentation performance with the number of learnable ISP parameters. For methods sharing the demosaicing frontend, the horizontal axis excludes that common frontend; the RAW-to-RGB methods are marked separately. This plot describes parameter cost, not end-to-end latency.

\subsection{Interpretation of Local Rendering Controls}

The local ISP keeps the reconstructed linear RGB fixed and changes only its rendered output. Its branches are parallel residual adjustments computed from the same base image, not a sequence in which tone first changes the input of the color or detail branch. Let $d(p)=\alpha_d\sum_i a_i(p)D_i(p)$. Since $\mathbf{w}^{\top}\mathbf{1}=1$ and $\mathbf{w}^{\top}\Delta_{\mathrm{color}}=0$, the pre-clipping brightness change is
\begin{equation}
    \Delta Y(p)=\alpha_t t(p)g(Y(p))+d(p).
\end{equation}
Thus, color adjustment lies in the brightness-neutral subspace defined by $\mathbf{w}$, whereas tone and detail change brightness with different spatial parameterizations. Tone uses smooth coefficient maps; detail modulates multiscale spatial bands. Clipping can alter this exact separation at saturated pixels. The resulting parameterization is interpretable, but task-level gains need not add because all residuals affect the same image and are optimized through the same downstream model.

\paragraph{Component results.}
All variants start from the same L Stage-1 checkpoint with the demosaicing frontend frozen. The fixed-ISP reference continues training only the downstream model, whereas the local-ISP variants jointly optimize the rendering controls and downstream model. All variants use the same training data and number of additional training epochs. After this training, the fixed-ISP reference achieves 80.023 IoU, 87.508 Dice, and 68.121 Boundary F1. The table below records the absolute values behind the plot.

\begin{table}[htbp]
\centering
\caption{Local ISP component results on COCO box-prompted segmentation.}
\label{tab:supp_local_components}
\setlength{\tabcolsep}{4pt}
\begin{tabular}{lrrr}
\toprule
Setting & IoU & Dice & Boundary F1 \\
\midrule
Fixed ISP (continued training) & 80.023 & 87.508 & 68.121 \\
Tone & 80.342 & 88.086 & 69.827 \\
Color & 80.311 & 88.052 & 69.955 \\
Detail & 80.200 & 87.970 & 69.806 \\
Tone + Color & 80.288 & 88.032 & 69.784 \\
Tone + Detail & 80.165 & 87.949 & 69.656 \\
Color + Detail & 80.281 & 88.017 & 69.983 \\
\midrule
Full & \textbf{81.152} & \textbf{88.640} & \textbf{71.570} \\
\bottomrule
\end{tabular}
\end{table}

\section{Limitations and Discussions}

The experiments establish task gains under the evaluated detection and box-prompted segmentation protocols. Fidelity retention depends on the frontend: L largely preserves its reconstruction scores, whereas S exhibits a stronger reconstruction--perception trade-off. The smaller model reduces parameter count, but parameter reduction alone does not establish device latency or energy savings. Real-capture evaluation covers 100 annotated images and measures class-agnostic prompted masks; extending camera coverage and evaluating automatic instance discovery are separate directions.

\end{document}